\documentclass[acmlarge]{acmart}

\usepackage[ansinew]{inputenc}
\usepackage{array}
\usepackage{makecell}
\usepackage{amsthm}
\usepackage{amsfonts}
\usepackage{color}
\usepackage{soul}
\usepackage{rotating}
\usepackage{xcolor,colortbl}
\usepackage{subfiles}
\usepackage{multicol}
\usepackage{stfloats}
\usepackage{url}
\usepackage{ulem}
\definecolor{Gray}{gray}{0.9}
\usepackage{balance}
\usepackage{xcolor}
\usepackage{amsmath}
\usepackage{tikz}
\usetikzlibrary{shapes}
\newcolumntype{C}[1]{>{\centering\let\newline\\\arraybackslash\hspace{0pt}}m{#1}}
\usepackage{rotating}
\usepackage{pdflscape}
\usepackage{afterpage}
\usepackage{capt-of}
\usepackage{longtable}
\usepackage{stfloats}
\usepackage{amsmath,amsfonts}
\usepackage{algorithmic}
\usepackage{graphicx}
\usepackage{caption}
\usepackage{textcomp}
\usepackage{longtable}
\usepackage{enumitem}
\usepackage{rotating}
\usepackage{subfigure}
\usepackage{hyperref}
\usepackage{amsxtra}
\usepackage{natbib}
\usepackage{subfigure}
\usepackage{latexsym}
\usepackage{caption}
\usepackage{amsmath,amsfonts,amsthm}
\usepackage[ruled,vlined]{algorithm2e}
\usepackage{multirow}
\definecolor{Gray}{gray}{0.9}
\usepackage{balance}
\usepackage{etoolbox}

\makeatletter
\def\ps@pprintTitle{%
   \let\@oddhead\@empty
   \let\@evenhead\@empty
   \def\@oddfoot{\reset@font\hfil\thepage\hfil}
   \let\@evenfoot\@oddfoot
}
\makeatother
\AtBeginDocument{%
  \providecommand\BibTeX{{%
    \normalfont B\kern-0.5em{\scshape i\kern-0.25em b}\kern-0.8em\TeX}}}

\setcopyright{acmcopyright}
\copyrightyear{2022}
\acmYear{2022}
\acmDOI{10.1145/1122445.1122456}

\acmJournal{CSUR}
\begin{document}
\title{XAI for Cybersecurity: State of the Art, Challenges, Open Issues and Future Directions}

\author{Gautam Srivastava}
\authornotemark[1]
\email{srivastavag@brandonu.ca}
\affiliation{%
  \institution{Department of Mathematics and Computer Science, Brandon University}
  \city{Brandon}
  \state{Manitoba}
  \country{Canada}
  \postcode{R7A 6A9}
}

\author{Rutvij H Jhaveri}
\authornotemark[2]
\email{rutvij.jhaveri@sot.pdpu.ac.in}
\affiliation{%
  \institution{Department of Computer Science and Engineering, School of Technology, Pandit Deendayal Energy University}
  \city{Gandhinagar}
  \state{Gujarat}
  \country{India}
  \postcode{382007}
}
\author{Sweta Bhattacharya}
\authornotemark[3]
\email{sweta.b@vit.ac.in}
\affiliation{%
  \institution{School of Information Technology, Vellore Institute of Technology}
  \city{Vellore}
  \state{Tamilnadu}
  \country{India}
  \postcode{632014}
}
\author{Sharnil Pandya}
\authornotemark[4]
\email{sharnil.pandya@sitpune.edu.in}
\affiliation{%
  \institution{Symbiosis Institute of Technology, Symbiosis International Deemed University}
  \city{Pune}
  \state{Maharashtra}
  \country{India}
  \postcode{412115}
}
\author{Rajeswari}
\authornotemark[3]
\email{rajeswari.c@vit.ac.in}
\affiliation{%
  \institution{School of Information Technology, Vellore Institute of Technology}
  \city{Vellore}
  \state{Tamilnadu}
  \country{India}
  \postcode{632014}
}
\author{Praveen Kumar Reddy Maddikunta}
\authornotemark[3]
\email{praveenkumarreddy@vit.ac.in}
\affiliation{%
 \institution{School of Information Technology, Vellore Institute of Technology}
  \city{Vellore}
  \state{Tamilnadu}
  \country{India}
  \postcode{632014}
}
\author{Gokul Yenduri}
\authornotemark[3]
\email{gokul.yenduri@vit.ac.in}
\affiliation{%
 \institution{School of Information Technology, Vellore Institute of Technology}
  \city{Vellore}
  \state{Tamilnadu}
  \country{India}
  \postcode{632014}
}
\author{Jon G.~Hall}
\authornotemark[5]
\email{Jon.Hall@open.ac.uk}
\affiliation{%
  \institution{School of Computing \& Communications, The Open University}
  \city{Kents Hill}
  \state{Milton Keynes }
  \country{United Kingdom}
  \postcode{MK7 6AA}
}
\author{Mamoun Alazab}
\authornotemark[6]
\email{alazab.m@ieee.org}
\affiliation{%
  \institution{Charles Darwin University}
  \city{Ellengowan Dr}
  \state{Casuarina NT}
  \country{Australia}
  \postcode{0810}
}
\author{Thippa~Reddy~Gadekallu}
\authornotemark[3]
\email{thippareddy.g@vit.ac.in}
\affiliation{%
  \institution{School of Information Technology, Vellore Institute of Technology}
  \city{Vellore}
  \state{Tamilnadu}
  \country{India}
  \postcode{632014}
}
\renewcommand{\shortauthors}{Gautam Srivastava, et al.}


\begin{abstract}
In the past few years, artificial intelligence (AI) techniques have been implemented in almost all verticals of human life. However, the results generated from the AI models often lag explainability. AI models often appear as a blackbox wherein developers are unable to explain or trace back the reasoning behind a specific decision. Explainable AI (XAI) is a rapid growing field of research which helps to extract information and also visualize the results generated with an optimum transparency. The present study provides and extensive review of the use of XAI in cybersecurity. Cybersecurity enables protection of systems, networks and programs from different types of attacks. The use of XAI has immense potential in predicting such attacks. The paper provides a brief overview on cybersecurity and the various forms of attack. Then the use of traditional AI techniques and its associated challenges are discussed which opens its doors towards use of XAI in various applications. The XAI implementations of various research projects and industry are also presented. Finally, the lessons learnt from these applications are highlighted which act as a guide for future scope of research. 

\end{abstract}



\maketitle

\begin{table}[!ht]
    \renewcommand{\arraystretch}{1}
    \caption{{List of key acronyms}}
    \centering
    \begin{tabular}{|p{1.5cm}|p{8cm}|}
        \hline
        \textbf{Acronyms} & \textbf{Description}\\
        \hline
        {AI}    & {Artificial intelligence} \\ \hline
        {ALARP} & {As Low As Reasonably Practicable} \\ \hline
        {ALE}   & {Accumulated Local Effects} \\ \hline
        {APA}   & {Advanced Persistent Attacks} \\ \hline
        {APTs}  & {Advanced Persistent Threats} \\ \hline
        
        {ATTCK}   & {adversarial Tactics, Techniques and Common Knowledge} \\ \hline
        {CF}    & {Counterfactual} \\ \hline
        {CIA}   & {Confidentiality, Integrity
                  and Availability}  \\ \hline
        {CNS}   & {Computer and Network Systems} \\ \hline
        {CPS}   & {Cyber Physical Systems}  \\ \hline
        
        {DDOS}  & {Distributed Denial-of-service } \\ \hline
        {DiCe}  & {Diverse Counterfactual explanation} \\ \hline
        {DNN}   & {Deep Neural Network} \\ \hline
        {DoS}   & {Denial-of-Service} \\ \hline
        {DT}    & {Decision Tree} \\ \hline

        {EU}   & {European Union} \\ \hline
        {FBI}   & {Federal Bureau of Investigation} \\ \hline
         {IAM}   & {Identity and Access Management} \\ \hline
         {IOC}   & {Indicator of Compromise}  \\ \hline
         {IDS}   & {Intrusion Detection System} \\ \hline
        {IoMT}  & {Internet-of-Medical-Things} \\ \hline
        {IoT}   & {Internet-of-Things} \\ \hline
        {ISO}  & {International Organization for Standardization} \\ \hline
        {ML}    & {Machine Learning} \\ \hline
        {MITM}  & {Man-in-the-Middle Attack} \\ \hline
        {NIST}  & {National Institute of Standards and Technology} \\ \hline
        {NSF}   & {National Science Foundation } \\ \hline
        {PMI}   & {Permutation Feature Importance} \\ \hline
        {QoS}   & {Quality-of-Services} \\ \hline
        {RPS}   & {Requests Per Second} \\ \hline
        {SDN}   & {Software-Defined Networks}\\ \hline
        
        {SHAP}  & {Shapley Additive explanations} \\ \hline
        {SIEM}  & {Security Information and Event Management} \\ \hline
        {SOC}   & {security Operation Centre} \\ \hline
        {SPATIAL}  & {Security an Privacy Accountable Technology Innovations, Algorithms and Machine Learning} \\ \hline
        {SVM}   & {Support Vector Machine} \\ \hline
        {VANET} & {Vehicular Ad-hoc Networks}\\ \hline
        {VNF}   & {Virtual Network Functions}\\ \hline
        {XAI}   & {Explainable Artificial intelligence} \\ \hline
        
    \end{tabular}
    \label{Tab:acronym}
\end{table}

\section{Introduction}
\label{intro}
Cybersecurity is the application of processes, controls and technologies to protect data, programs, network and systems from potential cyber attacks \cite{pienta2020can}. A variety of tools and technologies pertinent to cybersecurity aims at battling threats targeted towards network systems and applications existing in the internal or external environment of an organization. Statistical data reveal that the mean cost of  a data breach is 3.86 million dollars world-wide, rising to 8.64 million dollars in the USA \cite{mukhopadhyay2022cyber}. These costs include not only the direct impact of the breach but also the subsequent investigation to identify the reason for breaches, the associated responses associated, the loss of revenue, downtime and most importantly reputational brand damage\cite{krutilla2021benefits}. 

Mindful of these costs, most organizations have adopted cybersecurity strategies based on predominant best practices. 
An efficient cybersecurity strategy usually includes layered protection that provides defence against cyber attacks to preserve the confidentiality, integrity and availability of cyber assets. The implementation of such strategies also aims to prevent cases of financial extortion from users or reputed organizations which hinder normal business operations. Thus deployment of intelligent, effective and efficient counter measures in this regard is an absolute necessity. 
The National Institute of Standards and Technology (NIST), for instance, has developed a cybersecurity framework that helps organizations to protect their computer systems, networks and various other assets that are used for enabling national security, public health, safety and various other administrative activities. The International Standards Organisation, \textit{viz}., the ISO27000 family of information security standards aims to fulfill similar needs. Inspite of the existence of such approaches and standards, the attackers still find vulnerabilities in security frameworks wherein extremely strong defense measures get circumvented. During the pandemic crises, when the professional norms changed from in-office to work-from-home, cybersecurity threats also observed change in vulnerabilities relevant to remote access tools, cloud services and other remote working tools. \cite{pranggono2021covid}. Such evolving threats include Malware, Ransomware, Phishing, Insider Threats, Distributed Denial-of-service (DDOS) threats, Advanced persistent threats (APTs), Man-in-the-middle attacks and  various others \cite{natarajan2020cyber}. 

The cybersecurity frameworks and related best practices enable efficient reduction in cyber vulnerabilities by protecting confidential information without compromising on users' privacy and customer experience. To be more specific, Identity and access management (IAM), for instance, frames user roles and access rights, establishing criteria by which access privileges can be monitored. IAM technologies include single sign-on functionality, wherein users access a network without re-entering their credentials multiple times. IAM can also provides multi-factor authentication and privileged user accounts that provide access to specific legitimate users only, reducing the possibilities for fraudulent access. These types of tools provide enhanced visibility of anomalous activities in end-user devices. Also, in scenarios of security breach, the tools ensure accelerated investigation, response, isolation and containment of all the components pertinent to the security breach. 

There are various comprehensive data security platforms which include functionalities such as classification, permission analytics, behaviour analytics, and compliance reporting. The main objective of these platforms include protection of sensitive information in hybrid and multi-cloud environments. These platforms provide automated, real-time visibility, breach alerts and monitoring of data vulnerabilities \cite{callegati2016data}. As an example,  Security information and event management (SIEM) is a combination of security information management (SIM) and security event management (SEM) which involves the process of automating real-time analysis of security alerts generated by applications and network hardware. These products include intelligent and advanced detection methods, user behaviour analytics and artificial intelligence/machine intelligence (AI/ML) to detect anomalies\cite{azodi2013new} in the field of software products and services.

Cybersecurity risk management helps to understand the various characteristics of security threats and the relevant internal interactions at the individual and organizational level. The As Low As Reasonably Practicable (ALARP) is a similar risk management principle emphasizing on cyber risks. This principle ensures that the residual risks are reduced by comparing the a risk against the time, resources required in resolving the same. The idea is to analyse the the cost involved in reducing the risk and ensure it is not disproportionate to the benefit gained. All modern risk management solutions for cyber/information security are concerned with lowering the risk impact thereby balancing the cost associated in reduction or mitigation of the same. 

It is important to mention the scope of the International family of standard grouped as ISO27000 which emphasizes on the creation and management of documentations in Information Security Management Systems relevant to cybersecurity risks. The standard is organised around 114 controls in 14 groups and 35 control categories that cover all aspects of an organisation's cyber security. In order to apply the standard, one must assess the existing risks, identify the controls that apply, assess the mitigation that the controls render, assess the costs of applying those controls and also assess the mitigation of any secondary risks that are introduced. A control will be applied if 
\begin{enumerate}
\item the risk is assessed to be exceeding the risk appetite of the organisation;
\item the cost of the control application is deemed acceptable;
\item the secondary risks do not exclude an application.
\end{enumerate}
\subsection{How AI help can in Cybersecurity}

Machine learning (ML) algorithms are trained on the basis of previous experiences in order to take decisions resembling human behaviour. Additionally, ML algorithms have been used to detect anomalies and threats relevant to security threats and breaches\cite{almseidin2017evaluation}. In addition, automated ML-based security tools have been evolved in the past few years which respond automatically to threats, performing tasks such as clustering, classification and regression \cite{chan2019survey}. Clustering is a process wherein data are grouped based on their similarities of characteristics. The data objects in a cluster are similar to one another and different from the objects belonging to other clusters. Cluster analysis, thus, enables unsupervised classification of data having no predefined classes. Classification on the other hand helps to predict the class of given data points. A classifier uses training data to understand if the input variables belong to a particular class using unsupervised learning techniques. Regression analysis is a statistical technique that establishes the relationship between dependent and independent predictor variables with one of many independent variables.

AI and ML have also been applied for the purpose of pro-active vulnerability management. AI/ML-based User and Event Behaviour Analytics (UEBA) tools analyze user interactions at the service end points and at servers to  detect anomalous behaviour. This helps in providing prior protection to organizations long before vulnerabilities are reported or patched\cite{disterer2013iso}. 


Antivirus detection is an area where AI techniques have played a significant role. The most predominant approaches are heuristic techniques, data mining, agent techniques and artificial neural networks \cite{kamoun2020ai}. As an example, Cylance smart antivirus products are products, aligned to meet similar objectives providing enterprise level AI-based security to households detecting malware from legitimate data. The product eliminates the threat exactly at the point of execution without the need for any human intervention \cite{samtani2020cybersecurity}. 
There are various traditional authentication systems that use the username or email and password as a method of authentication. The use of AI helps in the detection of vulnerable password and is used in biometric-based authentication systems providing stronger layers of protection, which are difficult for hackers to compromise. The biometric systems are primarily used for enabling security and access control in businesses and government organizations. Biometric systems can be dichotomised into physical and behavioural  identification and recognition systems. The physical biometric systems use physical, measurable and distinctive information of the human body such as the DNA, veins, fingerprints, iris and others, and converts the same into understandable code by the AI system. On the contrary, the behavioural recognition systems capture unique behavioural characteristics such as voice, individuals typing rhythm, way of interaction with an object, and then stores such encoded information in a database. This information is digitally stamped during the authentication and verification process \cite{jaswal2021ai}. 

\subsection{Limitations of AI that necessitates XAI in Cybersecurity}

\begin{figure}[h!]
    \centering
	\includegraphics[width=\textwidth]{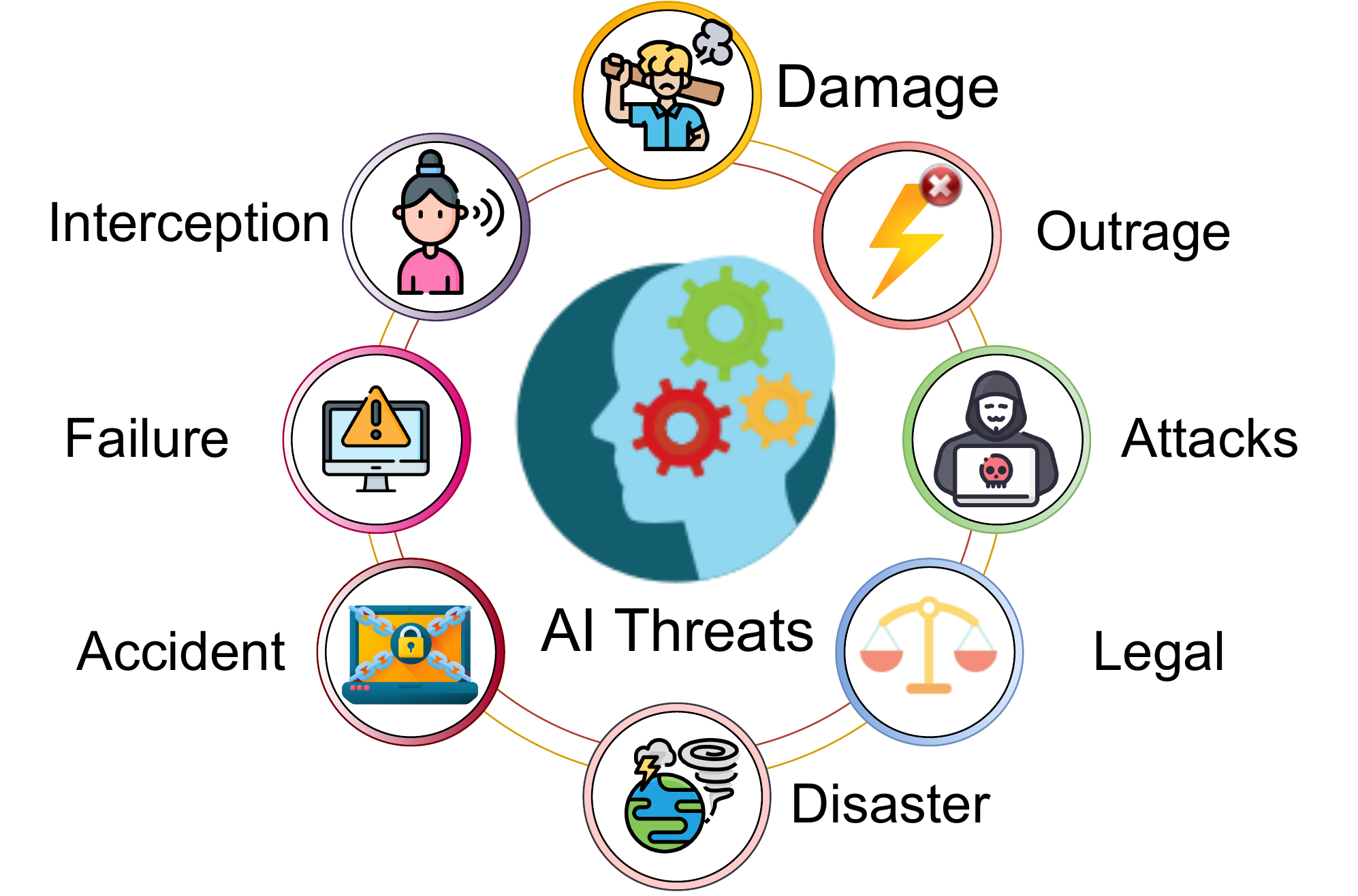}
    \caption{AI threat landscape}
    \label{fig:AI1}
\end{figure}

The applications of AI in cybersecurity induce many challenges. In particular, there are numerous counter-indications and secondary risks introduced by the applications of AI, which act as vectors for the attacks made by malicious actors. As an example, there could be instances wherein an attacker successfully avoids ML based detection. To be more specific, the attackers may manipulate the malware files such that the AI-based detection framework fails to identify any malicious or anomalous activity which is commonly known as evasion attacks. Similarly there exists various threats to AI-based cybersecurity applications pertinent to interception of communication, failure of services, accidents, disasters, legal issues, attacks, power outages and physical damages as portrayed in Figure \ref{fig:AI1}
 
The success of AI-based systems depends on the availability of data. AI-based systems raise two types of secondary risks. The first types includes the risk of generating false negatives results leading to inaccurate decision making. The second includes the risk of generating false positive results in which, there are possibilities of inaccurate notifications or false alarms.  \cite{nikolskaia2021relationship}. In such cases, there exists a dire need to ensure necessary mitigation steps are initiated to ensure scenarios of breaches or exceptional events are handled more accurately thereby maintaining interpretability and justifiability\cite{gerlings2020reviewing} of the decisions made.

Real-time AI systems often consume heavy computational power, data and raw memory resource requirements. These systems also require higher level of expertise to build and maintain \cite{alrajhi2020survey} making them immensely expensive to deploy. AI powered bio-metric systems have similar challenges and in association to the aforementioned issues these systems are also vulnerable to the risk of information breaches\cite{jaswal2021ai}. 

Cybersecurity firms predominantly use AI to develop robust and secure systems. On the contrary, these systems often get compromised for unethical purposes by hackers who train or mutate malware to become AI immune, behaving abnormally in comparison to traditional systems. The use of AI, enable hackers to foil the security algorithms making data manipulation remain undetected, thereby making it extremely difficult for organizations to rectify the data which are fed into AI-based security systems. 

Thus the current challenge with AI-based systems lies in the lack of rationale and justifiability of their decisions in comparison to the model-based traditional algorithms \cite{perarasi2020malicious}. If the systems fail to understand and learn from cybersecurity incidents then, no matter how powerful and accurate an AI-based system is, cybersecurity becomes a black box having pervasive secondary risk. 

In case of the deep reinforcement learning, the significant features which are identified as the reason for certain reactions, generally remain unexplained. As an example, the computation of Bayesian inference can be considered, wherein the accuracy of results generated often are impacted by issues pertaining to insufficient data. This initiates the need for statistical AI algorithm that help to quantify such uncertainties.  But the results of such statistical AI algorithms are often difficult to interpret, and thus, XAI plays its role by providing interpretability to the results generated by the AI based statistical models providing researchers and experts the ability to understand the causal reasoning and primary data evidence \cite{rudin2019stop}. Similarly, in case of the healthcare sector, the implementation of XAI firstly allows machines to analyze data and reach a conclusive result. Secondly, it enables doctors and other healthcare providers to get the decision lineage information explaining how a particular decision is reached. In the manufacturing industry, AI-based natural language processing (NLP) helps to analyze the unstructured data relevant to equipments and maintenance standards in association with structured data namely work orders, sensor readings and other business process data. This helps the technicians in reaching the best decisions in terms of their workflow related actions. 
\subsection{How XAI can help}
AI models have been successfully applied across many increasingly complex domains, complementing and augmenting human capabilities through their synthetic ability on the basis of complex dataset. The increase in computational power have further extended the scope of providing solutions through AI and the growth of AI applications have visualized exponential growth. As a consequence the demand for such AI applications in mission-critical settings have grown rapidly, wherein AI is being embedded in numerous hardware smart devices enabling unsupervised or remote-controlled usage. However, the applications of AI carry associated major issues. Overfitting, is one such fundamental issue in supervised ML wherein the statistical model fits perfectly with the training data hindering its ability to analyse accurately in cases of unknown data. The efficiency and accuracy of the model get decreased  as it captures noise  and inaccurate values in the data (Ying, 2019). The use of an over-fitted model leads to inferior performance of AI which in a mission-critical setting, may lead to inaccurate decision making, financial losses, physical harm or even death.

These risks can be mitigated to some extent by developing an understanding of the mechanism and reasoning of the model. Unfortunately, the black-box nature of the traditional AI systems acts as a bottleneck even for AI experts to provide justifiable solutions for the same \cite{storeyexplainable,shin2021effects}. Transparency is, thus, a necessity which would enable informed and justifiable decision making and also help in providing accurate explanation for the behavior of the models. As an example, in case of cybersecurity systems, unjustifiable and misleading predictions may make systems extremely vulnerable to attacks leading to completely unsecured critical systems. With the implementation of explainable AI, providing practical and real-time AI-based solutions would become much easier as biases in the datasets could be completely eliminated leading to impartial decision making. Interpretative results makes AI solutions much robust and trustworthy ensuring meaningful variable inference and basis of the models reasoning. The traditional deep neural network based models (DNN) are extremely popular but lag interpretability. As an example, in case of IDs, the reasoning behind intrusion detection becomes extremely difficult for network administrators to understand turning it into a black-box model. In such black-box models, the process involved in decision making is challenging to interpret as the DNN edits features on a trial and error to generate an ideal solution. Although numerous studies have been conducted on ML based IDS systems, very rarely the underlying reasoning or interpretation of the results are explored while reaching to conclusions relevant to classification of attacks, recognition of anomalous traffic behaviour and automated construction of the model. Decision Trees (DT) act as a perfect model to support explanations on the prediction of the results. The results of DT are not based on any assumptions relevant to distribution of the data and also handles feature collinearity issues efficiently. Thus, implementation of explainable AI systems enables network administrators analyze, explain and achieve insight on security policies in IDS systems\cite{mahbooba2021explainable,mane2021explaining}. 
In this paper, we explore the competing nature of cyber and AI risks, and explore the potential of XAI being implemented as the primary control against AI risk.
Numerous studies have been conducted considering XAI implementations in cybersecurity. Some of the studies are discussed in this section. The study by \cite{kuppa2021adversarial} proposed a novel black-box attack which implemented XAI to compromise the privacy and security of the relevant classifiers. The counterfactual explanation (CF) generation method was used in the study for implementing gradient-based optimization. The CF methods used in the study included latent CF technique, Diverse Counterfactual Explanations (DiCE) technique and permute attacks which performed end-to-end evasion attacks on anti-virus engines. They also performed membership inference attacks which helped to link users and hack their passwords from leaked dataset thereby launching poisoning and model extraction attacks on the same. The study evaluated the security threats pertaining to each of these attacks and provided information to users and attackers enabling scope of risk avoidance and mitigation. The study by \cite{marino2018adversarial} proposed a methodology for explaining inaccurate classifications generated by data oriented IDSs. Adversarial techniques were implemented to identify the minimum modification in the input attributes which were required to accurately classify the misclassified samples of the dataset. In \cite{mane2021explaining}, a deep learning based intrusion detection framework was proposed. The explainable AI techniques in the study, helped to achieve transparency at each level of the ML model.
\begin{figure}[h!]
    \centering
	\includegraphics[width=\textwidth]{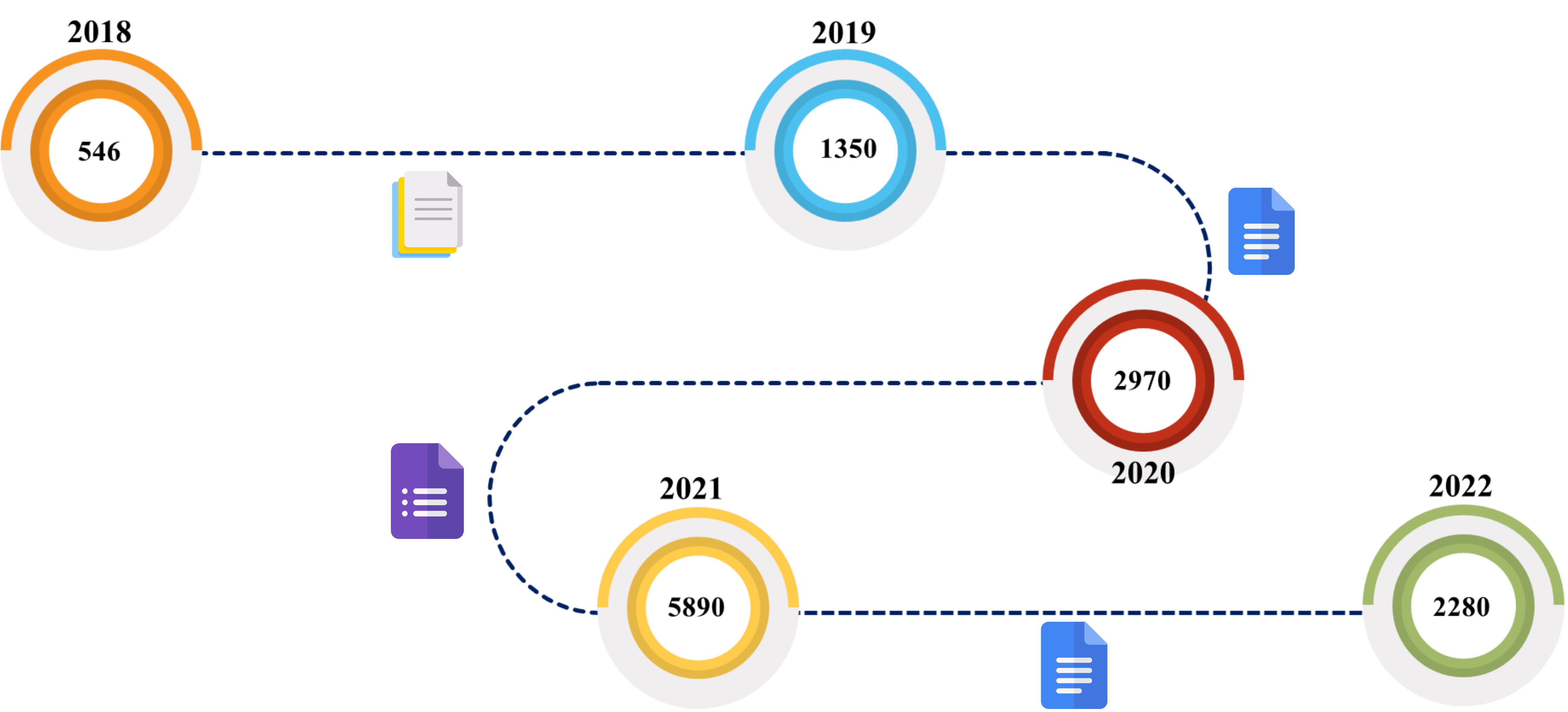}
    \caption{Recent number of works on XAI as per Google scholar statistics.}
    \label{fig:statics}
\end{figure}

XAI has made significant progress in various industry verticals, including banking, healthcare, manufacturing, and defence. XAI enables stakeholders to better comprehend and interpret an AI model's predictions. Since 2018, 546 researchers have adopted XAI in their research, and this number has kept increasing to 1,350, 2,970, 5,890, and 2,280 in the years 2019, 2020, 2021, and 2022 respectively, according to Google scholar statistics as on 10$^{th}$ of May, 2022. Fig.~\ref{fig:statics} explains the recent number of works on XAI as per Google scholar statistics.

The XAI methods used in the study included SHAP, BRCG which enabled complete understanding of the models behaviour. The SHAP and CHEM techniques of XAI helped to understand features of the input which led to a decision derived as an output. Considering the analysts perspective, Protodash method was used for the identification of differences and similarities between training dataset samples. The authors in \cite{rao2021zero} proposed an innovative approach to manage over-flooding issues in alarm systems of cybersecurity systems. The systems considered for implementation in the study included security information and event management (SIEM) system and intrusion detection system (IDS). A zero-shot learning technique was integrated with ML for computing the explanations of the anomaly predictions in the framework. The unique approach of the framework involved identification of the attack without any prior knowledge, deciphering of the features that lead to classification  and then grouping the attacks in a particular category using XAI technique. The use of XAI helped to identify, quantify factors and understand its contribution towards particular prediction of cyber attacks. The study in \cite{mahbooba2021explainable} proposed an enhanced trust management system in IDS using a decision tree based XAI model. The decision tree algorithms used in the study helped to split the choices in multiple sub-choices for IDS, thereby generating rules for benchmark dataset. The results of the decision tree based XAI approach yielded enhanced accuracy than the traditional Support Vector Machine (SVM) system. 

\begin{figure*}[h!]
\centering
\includegraphics[width=\textwidth]{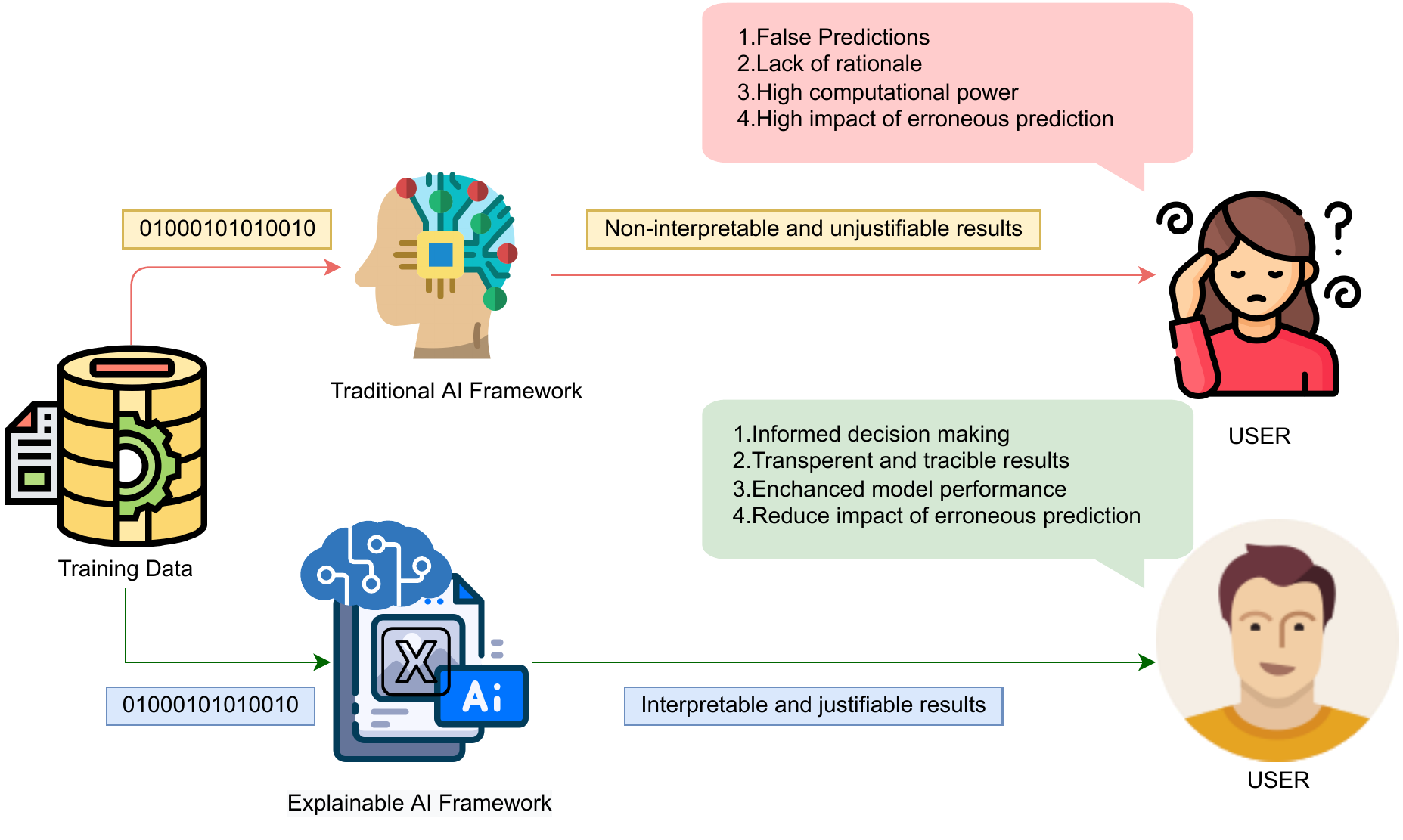}
\caption{\textcolor{black}{AI and XAI-user's perspective}}
\label{fig:AIvsXAI}
\end{figure*}

Although there exist various review articles focusing on applications of AI in cybersecurity, but a comprehensive review of the application of explainable AI in cybersecurity is not available yet, which includes explicit and extensive information. Hence, to bridge this gap, the present paper emphasizes on providing an extensive review of XAI in cyberseurity highlighting the existing state of the art, the challenges involved in existing AI implementation, the need for XAI and its potential scope of applications in various sectors. A comparative analysis of the existing work on XAI and the present paper is highlighted in Table \ref{tab:Summary }. The benefits of using XAI over AI from a users perspective is highlighted in Figure \ref{fig:AIvsXAI}. 
To summarize, the specific contributions of this study includes:

\begin{itemize}
    \item Basic information on Cybersecurity and various forms of attacks. 
    \item The need for XAI implementation is justified highlighting various applications of AI in cybersecurity and related lacuna in providing explanation of the generated results
    \item The applications of XAI-based cybersecurity frameworks in various industries are presented
    \item Various research projects and industry projects implemented using XAI for cybersecurity are discussed in details
    \item The lessons learnt from these implementations are identified which help in identifying the future scope of the research in this area
\end{itemize}

 
 \begin{table*}[h!]
\centering
\caption{{Summary of literature survey}}
\label{tab:Summary }
\resizebox{\textwidth}{!}{%
\begin{tabular}{|l|l|l|l|l|l|p{12cm}|}
\hline
Ref. & {\rotatebox[origin=c]{90}{~AI~}} & {\rotatebox[origin=c]{90}{~XAI~}} & {\rotatebox[origin=c]{90}{~Technical Aspect~}} & {\rotatebox[origin=c]{90}{~Applications~}}& {\rotatebox[origin=c]{90}{~Future Challenges~}} & Remarks \\ \hline
[21] & \cellcolor[HTML]{FFFFC7}L & \cellcolor[HTML]{d1f8d1}H & \cellcolor[HTML]{FFCCC9}M & \cellcolor[HTML]{FFCCC9}M & \cellcolor[HTML]{FFFFC7}L & Implementation of decision tree algorithms based XAI approach in IDS using KDD dataset. \\ \hline
[22] & \cellcolor[HTML]{FFCCC9}M& \cellcolor[HTML]{d1f8d1}H & \cellcolor[HTML]{d1f8d1}H & \cellcolor[HTML]{d1f8d1}H & \cellcolor[HTML]{FFFFC7}L & Implemented DNN and XAI models for predicting intrusion detection. \\ \hline
[23] & \cellcolor[HTML]{FFFFC7}L& \cellcolor[HTML]{d1f8d1}H & \cellcolor[HTML]{d1f8d1}H& \cellcolor[HTML]{FFFFC7}L & \cellcolor[HTML]{FFFFC7}L & Emphasises XAI methods focusing on confidentiality and privacy of various classifiers.  \\ \hline
[24] & \cellcolor[HTML]{FFFFC7}L & \cellcolor[HTML]{FFCCC9}M & \cellcolor[HTML]{FFCCC9}M& \cellcolor[HTML]{FFCCC9}M& \cellcolor[HTML]{FFFFC7}L & Study generating rationale for inaccurate classifications of data driven IDS using XAI. \\ \hline
[25] & \cellcolor[HTML]{FFFFC7}L& \cellcolor[HTML]{FFCCC9}M & \cellcolor[HTML]{d1f8d1}H& \cellcolor[HTML]{FFFFC7}L & \cellcolor[HTML]{FFFFC7}L & Study of zero-shot approach for adapted cybersecurity for XAI. \\ \hline
Proposed & \cellcolor[HTML]{d1f8d1}H & \cellcolor[HTML]{d1f8d1}H & \cellcolor[HTML]{d1f8d1}H & \cellcolor[HTML]{d1f8d1}H & \cellcolor[HTML]{d1f8d1}H  & A comprehensive survey of XAI in cybersecurity including frameworks, application, research projects, real-time usecases, lessons learnt, bottle necks, and potential future direction of research.\\ \hline
\end{tabular}%
}
\begin{flushleft}
\begin{center}
    
\begin{tikzpicture}

\node (rect) at (0,2) [draw,thick,minimum width=1cm,minimum height=0.7cm, fill= red!25, label=0:Low Coverage] {M};
\node (rect) at (3.4,2) [draw,thick,minimum width=1cm,minimum height=0.7cm, fill= yellow!35, label=0:Medium Coverage] {L};
\node (rect) at (7.5,2) [draw,thick,minimum width=1cm,minimum height=0.7cm, fill= green!20, label=0:High Coverage] {H};
\node (rect) at (11,2) [draw,thick,minimum width=1cm,minimum height=0.7cm, fill= red!75, label=0:Not Applicable] {NA};

\end{tikzpicture}
\end{center}

\end{flushleft}

\end{table*}       

\section{Background and Motivation}
 This section discusses about the basic concepts of cybersecurity and predominant cyberattacks. The description of the attacks lay the foundation of the motivation behind implementing XAI in cybersecurity.
\subsection{Detailed Discussion on Cybersecurity}
According to the International Organization for Standardization (ISO/IEC 27032), cybersecurity is defined as the protection of digital information in terms of its confidentiality, integrity and accessibility.  ISO/IEC 27032 \cite{XAI} has recommended a policy framework for stakeholders on the internet to help develop trust, communicate with one another, exchange information and provide technical assistance for system integration. 

People, software and technology services from all over the world can communicate with one another through the internet. In parallel with the growing reliance of the international economy on cyberspace for its operations, an increasing number of cyberattacks are becoming increasingly widespread. Everyone in every organization, regardless of its size or industry, are seeking to develop a cybersecurity strategy which is both successful and comprehensive in its approach. The availability of the internet grew in popularity, and an increasing number of people began posting their personal information on the internet. In recognition of this, as a potential source of data, organized crime groups started exploiting the internet to steal personal information from individuals and governments. As a result, the requirement of cybersecurity has grown at an exponential rate, necessitating the widespread development of firewalls and antivirus software. Thereby, there is a tremendous amount of growth in the cybersecurity market. The global cybersecurity industry is expected to reach 345.4 billion USD by 2026 \cite{XAI1}.

\subsection{Cyberattacks}
\begin{figure*}[h!]
\centering
\includegraphics[width=\textwidth]{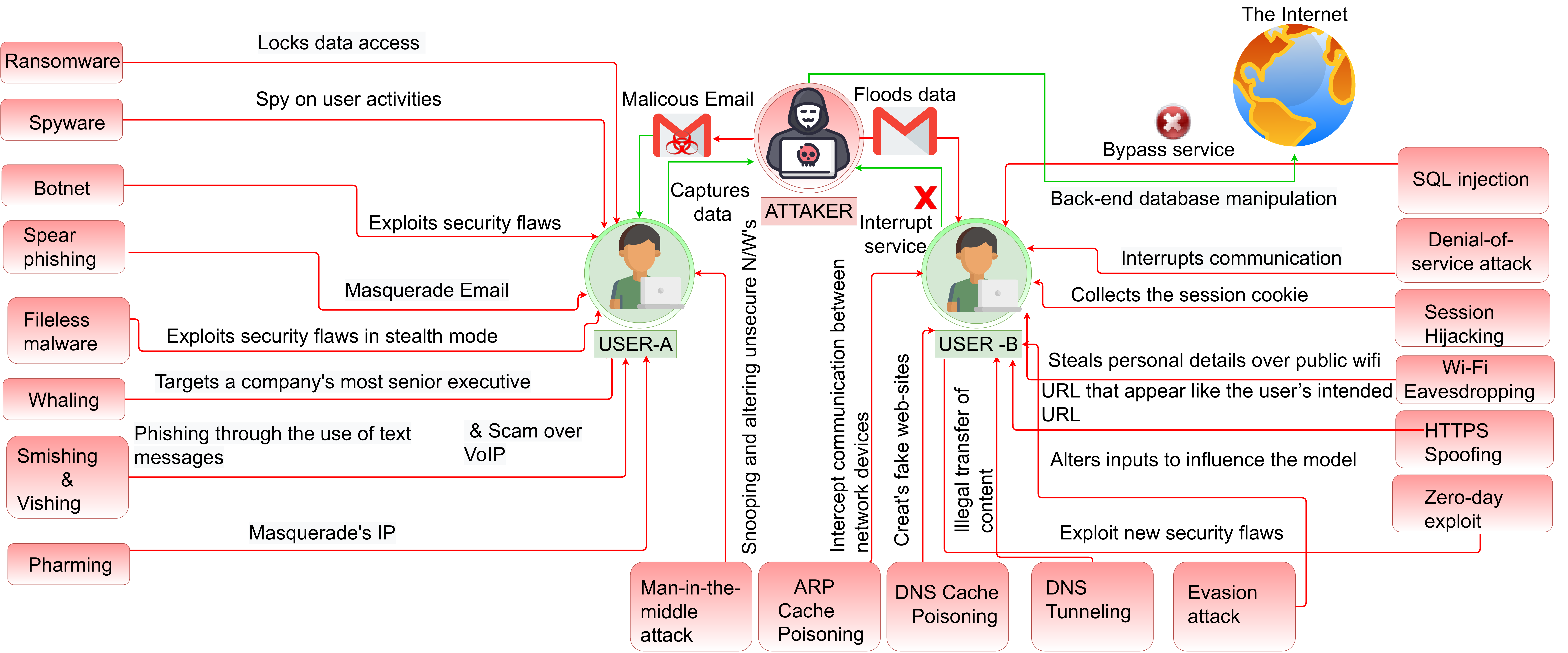}
\caption{Types of cyber attacks}
\label{fig:Attacks}
\end{figure*}

As our society has evolved, being more interconnected, cyberattacks have become even more sophisticated. Therefore, as data breaches grow more widespread, it is critical to have a thorough grasp of modern cyberattacks. This section provides an overview of some of the most prominent cyberattacks and is also depicted in Figure \ref{fig:Attacks}.
\subsubsection{Malware}
The term 'malware' refers to malicious software such as spyware, ransomware, viruses and worms. When an individual clicks on a bad link or opens an email attachment mistakenly, the machine may become infected with a malware. Once the malware gains access to a computer system, it can restrict access to key network resources. In the absence of user's knowledge, malware has the ability to collect and transmit sensitive information. Malware can potentially disable a large number of system components, making the machine utterly useless \cite{9061034}.

The following are some of the popular types of malware:
\begin{enumerate}
    \item Ransomware : It is a type of malware that demands payment and is considered to be one of the most severe types of malware. It encrypts the data and demands payment to decrypt it. One of the most prevalent reasons for this is that people unintentionally infect their machines with this type of virus through email attachments or links from untrustworthy sites. If ransomware is installed, it may provide a backdoor for hackers who may then encrypt the data on a device and prohibit its owner from accessing it until they pay a ransom. It is also referred to as crypto-malware because it demands ransom payments in digital money. To summarise, ransomware can take over computers, encrypt data and cause financial harm to the victim \cite{9371288}. The exploitation of Kaseya's Virtual System Administrator enables the REvil hacking group to spread false software update which compromises both Kaseya's direct clients and their customers. According to reports, one million systems have been seized demanding a ransom from REvil. Moreover, around 500 of Kayesa's clients and more than 1,000 enterprises have been impacted severely. In exchange for their services, the hackers sought USD 70 million in bitcoin mode. According to the firm, Coop, a Swedish grocery operator, was forced to close 800 of its stores for a whole week as a result of the cyber attack \cite{XAI2}. The FBI gained access to REvil's systems shortly after the attack and recovered the encryption keys required to resolve the breach's compromise. Fortunately, Kaseya's clients IT infrastructure was not compromised, and no ransom was paid. This could be considered as one of the most damaging ransomware attack of the year 2021.
    
    \item Spyware: It is a type of malware which infiltrates systems without the owner's permission. This is frequently done with the intent of spying on the internet activity, tracking login and password information or gathering sensitive information which could be used fraudulently. Spyware is a broad phrase that refers to a variety of unwanted software, such as adware, Trojan malware and even cookie tracker. One of the most popular types of spying software is known as "keylogger" which records every keystroke made on the keyboard and saves the data we enter. To summarise, spyware has the potential to violate users' privacy and, collect sensitive data, steal data or steal identity \cite{9013026}. WhatsApp, which is owned by Meta, previously called Facebook, confirmed for the first time in 2019 that the Pegasus spyware software was used by attackers to spy on 1,400 people in India, including journalists and human rights activists \cite{XAI3}.
    
    \item Botnet: A botnet is a "spider" program which monitors the internet for security flaws and prevents exploitation. The intrusion is carried out automatically in this attack. It infects devices by injecting them with harmful code. Botnets can be used to effectively hack devices. They are capable of launching distributed denial-of-service (DDoS) attacks and recording actions such as keystrokes, webcam footage or screenshots. A botnet can be used to get remote access to your computer by hackers. A 12-month study by Imperva Research Labs found that 57 percent of all attacks on e-commerce websites in 2021 were done by botnets, compared to 33 percent for all other industries \cite{9435800}. Cloudfare, a provider of internet infrastructure, was hit by a DDoS attack in early 2021 which was nearly three times as massive as any other DDoS attack they had seen before \cite{cloudfare}. More than 17 million fake traffic requests per second (RPS) were discovered by Cloudfare. It was a botnet called Meris that initiated the attack on a financial industry customer. Cloudfare received more than 332 million attacks from Meris because of this. More than 20,000 bots from 125 countries were responsible for the attack traffic, according to Cloudfare.
    
    \item Fileless malware: The term "fileless malware" refers to the malware which is installed and executed via software, programs and protocols which are already built into or native to the operating system of a device. The fileless malware does not require downloading of a file. It is a malware which does not make use of files as it is memory-based. As long as legitimate scripts continue to run, fileless malware can continue to cause havoc. Due to the stealthy nature of fileless malware, it is difficult to detect. As a result, fileless malware is capable of disrupting antivirus software and stealing \cite{9565863}. Astaroth fileless malware was a legitimate data-stealing malware in 2019, and it primarily affected Windows-based systems. If users do not exercise caution when downloading and running this program, they would be confronted by a flood of legitimate Windows tools, one after the other, in rapid succession. In fileless execution, all  programs download extra code and transfer their output to one another without saving any files on the disc, making the job of the traditional antivirus solutions more difficult as they would have no files to scan. The Astaroth malware is capable of dumping credentials for a broad range of applications and uploading the stolen information to a remote server \cite{Astaroth}.
\end{enumerate}

\subsubsection{Phishing}
Phishing is the term used to describe fraudulent email messages which appear to come from a well-known company. The goal is to steal sensitive information from the victim's computer, such as credit card and login information, or infect the victim's computer with malicious software. Phishing is a type of internet fraud which is becoming increasingly popular. There are a variety of phishing attacks \cite{8924660}.
\begin{enumerate}
\item Spear phishing: It is an email or any such electronic communication scam wherein the attacker tries to get financial benefits from a person, a business or an organization. To get access to the victim's data, attackers spread malicious software on the victim's computer. In the year 2015, Verizon estimated that almost 80 percent of all malware attacks are the result of phishing attempts using various forms of social engineering techniques. A spear-phishing campaign on Ubiquiti Networks Inc, which is a provider of network technology for service providers and enterprises in the United States, caused huge devastation \cite{6289402}. In June of 2015, the firm suffered a USD 46.7 million loss as a result of a spear-phishing email. A combination of employee fraud and bogus requests from an outside source was used to launch the attack against the Company's finance department. It is due to this scam, funds totaling USD 46.7 million were transferred from a company subsidiary incorporated in Hong Kong to other offshore accounts controlled by third persons, totaling USD 46.7 million. Employees at Ubiquiti were duped into believing that they were getting legitimate requests from the company's management as a consequence of the use of spoof e-mail addresses and look-alike domains. Although no sensitive information was stolen in this attack, the incident acted as a warning of how easily spear phishers can fool their targets into taking action by impersonating legitimate people and leveraging publicly available information to create convincing fake emails that look authentic \cite{Ubiquiti}.

\item Whaling: When a cyber attack targets a company's most senior executive, typically the CEO or CFO, this is referred to as whaling. Attackers impersonate senior or other key members of an organization and, directly target senior or other important persons. The goal is to steal money or confidential information from them or to gain access to their computer systems in order to commit cyber crimes \cite{ravi2020performance}. When an Australian hedge fund's co-founder clicked on a bogus Zoom link in November 2020, the malware was planted on their network. Fraudulent invoices were used to attempt stealing of USD 8.7 million from the assailants. A total of USD 800,000 were stolen from the stake holders. The hedge fund was forced to shut down as a result of losing its largest client. Aerospace company dismissed the CEO of the organization after USD 58 million whaling loss in another instance \cite{Hedge}. 

\item Smishing and Vishing: The term "SMS phishing" or 'Smishing' refers to the practice of phishing through the use of text messages. Those who are the victim of a Smishing attack are inundated with text messages that appear to come from a reputable source. Voice phishing, often known as Vishing, is a type of phishing attack which takes place over the phone \cite{jain2021survey}. The use of Voice over IP (VoIP) technologies by phishing scammers to interact with their victims is becoming increasingly common. Tripwire detected a Smishing campaign disguised as the US Postal Service. Senders of the malicious SMS messages instructed their targets to click on a link to obtain details about a forthcoming US Postal Service delivery. The malicious link directed consumers to a series of websites which were specifically tailored to steal their Google account information. According to the Spectrum Health systems in 2020, a Vishing assault, in which patients were contacted by someone claiming to be Spectrum Health System worker, was reported \cite{Spectrum}. Patient's and Spectrum Health members' personal information, including their member ID numbers and other sensitive health information, were the primary goal of the attackers. Attackers reportedly used charm and even threats to persuade victims into turning over personal information, money or access to their electronic devices.

\item Pharming: Attackers reroute the internet visitors wanting to reach a specific website to a fake one using the social engineering tactic known as 'Pharming'. Malware may be installed on the victim's computer through spoofed sites which seek to steal their personal information and log-in credentials. Pharmers frequently target financial institutions' websites such as banks, payment systems or e-commerce sites to steal personal information \cite{vayansky2018phishing}.   Pharming attacks were carried out in Venezuela during the first half of 2019 which were particularly noteworthy. The President announced the formation of "Voluntarios por Venezuela"--Voluntaries for Venezuela. The purpose of this initiative was to connect volunteers with humanitarian aid organizations. It obtained all of the information on the volunteers through a website. This included their entire name, ID number, phone number and location. The debut of the website was followed by the appearance of a virtually identical website with a similar domain name a few days later. The hacker used the same IP address as the original website to set up this fake domain in his or her name. Whenever a volunteer accessed the legitimate website, all personal information they supplied was filtered to the fake site, resulting in the data theft of thousands of volunteers from the legitimate website \cite{venezuela}.

\end{enumerate}

\subsubsection{Man-in-the-middle attack}
Attackers disrupt an ongoing communication or data transmission by placing themselves in the middle of it via a "man in the middle" (MITM) attack. Following their insertion, the attacker will claim himself legal to parties in the transfer. This allows an attacker to intercept the data and information from both the parties while also providing malicious links or other information to both genuine participants in a way that may not be detected immediately \cite{8949690}. There are a variety of Man-in-the-middle attacks.

\begin{enumerate}
\item ARP Cache Poisoning: ARP poisoning attack, also known as ARP spoofing attack, is a type of MITM attack which allows attackers to intercept communication between network devices. It has possibilities of poisoning the MAC of other devices to IP mappings using ARP by utilizing the protocols flaws. A malicious attacker can poison the ARP caches of other hosts on a local network by using readily available tools, causing the ARP caches of other hosts to be filled with erroneous entries \cite{7904631}.

\item DNS Cache Poisoning: It is a very deceptive cyberattack which redirects web traffic to phishing websites by poisoning the DNS cache, also known as DNS cache poisoning or DNS spoofing. Hackers can easily fool users into handing over personal information by creating fake websites which appear like the user's intended destination \cite{alharbi2022dns}.

\item HTTPS Spoofing: In HTTPS Spoofing, URL of an attacker's HTTPS site slightly differ from the URL of a legal site having valid authentication certificate. This allows hackers to steal personal information by creating a slightly different URL which appears like the user's intended URL \cite{alharbi2022dns}.

\item Wi-Fi Eavesdropping: This type of attack involves attackers snooping on Wi-Fi data on unsecured networks or constructing their networks with popular names to fool people into connecting, so that they can steal their login credentials or credit card details or anything else they send over that network \cite{vashist2019securing}.

\item Session Hijacking:Session hijacking is a type of MITM attack, which occurs when an attacker waits for the user to log into a web page and then, collects the session cookie and uses it to log into the same account from his browser \cite{hu2019session}.

\end{enumerate}

\subsubsection{Denial-of-service attack}
Denial-of-service (DoS) attacks are malicious and targeted attacks which can overwhelm a network with fraudulent requests in order to disrupt business activities. In sucg attacks the users are prevented from accessing resources which are stored on a system or in a network \cite{huang2022learning}. Despite the fact that the vast majority of DoS attacks do not result in data loss and are frequently handled without the payment of a ransom, they have a considerable impact on an organization's critical business processes in terms of resources and time.

\subsubsection{SQL injection}
Malicious SQL code is used for back-end database manipulation to get access to information which is not meant to display. This malicious code injection technique is called SQL injection--SQLI. The details such as confidential company information, user lists or private consumer information can be compromised using this attack \cite{li2019lstm}.

\subsubsection{Zero-day exploit}
The term 'zero-day' refers to newly found security flaws which can be exploited by the hackers. When a vendor or a developer discovers a vulnerability recently, it is referred to as zero-day vulnerability. A zero-day attack occurs when a vulnerability is exploited by hackers before it can be fixed by the software developers \cite{jin2022exgen}.

\subsubsection{DNS Tunneling}
Non-DNS traffic can be communicated via port 53 using DNS tunneling. Other protocols are also sent by DNS. DNS tunneling can be used for a variety of good reasons. However, DNS Tunneling VPN services can also be used for malicious purposes. Outgoing traffic disguised as DNS can be used to hide data which is routinely shared over an internet connection using these services. DNS requests can be modified to send data from a hacked machine to the attacker's infrastructure for malicious purposes. The attacker's infrastructure can also use it with the call-back commands to a compromised system \cite{9426926}.

\subsubsection{Evasion attacks}
Evasion attacks alter input data in order to evade the utility of an existing model, whereas poisoning attacks contaminate training data in order to contaminate a newborn model. Additionally, these attacks might be classified as white-box and black-box. A white-box assault is conducted on the assumption that complete knowledge of the target model, including its characteristics, architecture, training approach, and training data, is provided. A black-box assault is one that is conducted under the assumption that no information about the target model is available. Whether the adversarial attack is white-box or black-box, at the very least one model should be directly targeted, either the target model or a substitute model \cite{hao2022adversarial}.   

\subsubsection{Poisoning attack}
In poisoning attack, inputs are modified during training period and model is trained on modified inputs to obtain desired output \cite{9431105}. Poisoning attacks can be classified as random attacks or targeted attacks, depending on the attacker's intent. Random assaults seek to lower the model's accuracy, whereas targeted attacks seek to get the model to output the adversary-specified target label. In general, targeted attacks are more difficult to execute than random attacks, as the attacker is pursuing a defined objective. Poisoning attacks can be conducted on either the data or the model during the training process. Poisoned updates can originate from two types of poisoning attacks they are data poisoning during local data collection and model poisoning during local model training. At the higher level, both poisoning attacks aim to alter the target model's behaviour in an unwanted way. If adversaries access the server, they can easily conduct poisoning attacks on the learned model, both targeted and untargeted. If they compromise the server, they can simply perform both targeted and untargeted poisoning attacks on the trained model.

\subsection{Detailed Discussion on XAI}
In 2004, Van Lent et. al \cite{van2004explainable} initiated the term XAI to describe the ability of their technology to explain the behavior of AI-controlled entities in simulation-based gaming applications. The term XAI has no universally agreed-upon technical definition. The goal of XAI is to make the results of AI easier to understand by end-users.
DARPA \cite{DARPA} states that the goal of XAI is to create more explainable models while allowing stakeholders to better understand and appropriately trust the emerging generation of artificially intelligent partners. FICO \cite{FICO}, consider XAI as "an innovation towards opening up the black-box of ML" and "a challenge to produce models and procedures which are both accurate and provide a good trustworthy explanation which will satisfy customer's needs".

Expert systems have been the subject of explanation studies since long before this word was coined. There was a noticeable slowdown in the rate of progress toward tackling this problem after the tremendous gains in ML were made. As a result, the focus of AI research has turned in to creating models and algorithms which emphasize predictive capability, while the ability to explain decision processes has taken a sideline. Researchers and practitioners have recently given attention to XAI. According to the Google Trends, the interest in the XAI phrase has risen dramatically over the past few years \cite{adadi2018peeking}. An increasing interest in this study area is a direct result of AI/relentless ML's spread throughout sectors and its essential impact on vital decision-making processes, without being able to provide precise information about the chain of reasoning which leads to specific judgments. Because of this, new AI strategies are needed which are capable of making decisions which are intelligible and comprehensible to the general public.

 In summary, XAI is a collection of procedures and methods which helps researchers to understand and rely on the results and conclusions of a ML model. XAI aids researchers in figuring out how accurate, reasonable, transparent and effective AI-powered decisions are, by comparing them to other decisions. The outputs of AI models must be explainable to trust them while making real-world adaptations.

\subsection{XAI Frameworks}
In the field of AI and ML, the term "explainable AI" is gaining traction. Humans need to be able to put their complete trust in the AI models which make their decisions. Only by making the black-box of ML more visible can this be achieved. An XAI framework is a tool which generates reports on the model's operation and makes an attempt to explain how it operates. This section provides an overview of some of the most popular XAI Frameworks.

\subsubsection{SHAP}
SHapley Additive exPlanations (SHAP) is a framework for justifying and explaining the result of prediction models \cite{9141407}. Game theory is used to demonstrate a relationship between optimal credit distribution and local explanations using traditional Shapley values. The bar-length and color-coding approach used by SHAP makes it easier to understand the size and direction of an impact. It can be used to explain simple ML algorithms such as linear regression, logistic regression and tree-based models. Moreover, it can be used to explain more complex models such as deep learning models for image classification and captioning, and various natural language processing tasks such as sentiment analysis, translation and text summarization.
\subsubsection{LIME}
Considering the aspect of explainability, LIME was one of the first methods to provide the same \cite{9521165}. Local Interpretable Model agnostic Explanations (LIME) is much similar to SHAP, but it works much faster. LIME generates a list of explanations, with each explanation indicating the role of each feature in the forecast of a particular data sample. There is no black-box classifier which LIME cannot explain. The classifier implements a function which takes in raw text and outputs a probability for each class. Predictions for tabular data, images and text classifiers are currently explained by LIME.
\subsubsection{ELI5}
ELI5 framework is an MIT explainability package which supports ML \cite{9326354}. Many ML frameworks are supported by ELIS such as sci-kit-learn, Keras, XGboost, LightGBM and Catboost. The comparison of models across frameworks and packages becomes easier with the help of an unified API. Classification models and regression models can be interpreted in two ways. The first approach is to analyze the model parameters and try to understand the working of the model in its entirety. The second approach includes examination of an individual prediction in order to understand the reasoning behind a particular decision made by the model.
\subsubsection{Skater}
Skater is a model-agnostic framework for performing model interpretation on a variety of models \cite{arya2019one}. The framework is necessary when developing an interpretable ML system for real-world use cases. Deconstructing the learned structures of the black box model can be done globally or locally using this open source python library.

\subsubsection{DALEX}
The DALEX package allows researchers to better understand how the model behaves \cite{biecek2018dalex}. The most often used models are classification and regression in ML. Models such as boosting, bagging, stacking and other related methods are frequently employed. But these black-box models are difficult to interpret. The DALEX package includes a variety of methods for analyzing the relationship between model input and model output. The implemented approaches aid in the exploration of the model from a single instance to a complete dataset.

\subsubsection{ALE}
Accumulated Local Effects (ALE) is a global explanation technique that examines the relationship between feature values and target variables. ALE averages and adds the difference in predictions throughout the conditional distribution, thereby isolating the impacts of each feature. This comes at the cost of a greater number of observations and a nearly uniform distribution.In summary, it shows the main effects of individual predictor variables and their second-order interactions in black-box supervised learning models that are easy to understand. The package can make ALE plots based on fitted supervised learning model

\subsection{Motivation to Integrate XAI for Cybersecurity}
AI's potential to predict the future has attracted growing number of individuals in versatile verticals. Inspite of their extensive use, the outcomes of these models are difficult to defend. For example, thousands of parameters must be fine-tuned in the AI/ML model which can run a self-driving car. It is often difficult for an administrator to understand the AI/ML system's logic in the event of a security breach. Also,making decisions in such black-box environment is also tough. AI systems sometimes provide inaccurate findings in the form of false positives which mislead security experts jeopardizing the integrity of the entire system. Cybersecurity is one of the numerous sectors in which XAI can be beneficial because the model's outcomes would become explainable. Traditional AI and ML-driven approaches are no match for XAI's capacity to keep up with continually evolving threats, automatic threat detection, swift and efficient response capability. XAI provides security professionals the quality of essential reasoning and threat identification in order to reduce breach risk and improve overall security. XAI also has the scope to help with risk discovery and prioritizing, incident response coordination and malware threat detection. XAI appears to be a promising solution in circumstances requiring explainability, interpretability and accountability.

\subsubsection{Explainability}
The growth of the internet has also raised the chance of a network being compromised. Intruders are on a constant search for novel ways to breach the security of a network. Due to the black-box nature of AI-based security frameworks, they are unable to provide detailed information about identified attacks or reports which mentions ways of compromising a system. In this scenario, XAI appears to be a viable solution for explaining the detected intrusion while also revealing the attack's unique characteristics. Additionally, XAI can provide a report explaining the type of attack which can support cybersecurity experts in developing effective cybersecurity frameworks.

\subsubsection{Interpretability}
Cybersecurity models based on conventional AI techniques cannot provide clear insights about vulnerabilities. The results are unreadable which causes concerns among cyber domain users. In the absence of prior knowledge of ML models, the results become hard to interpret. The substitution of XAI in the place of traditional AI-based frameworks enables clear visualization of the report providing users the ability to comprehend potential damage caused by vulnerabilities. This would assist the users to be prepared for the future risks.

\subsubsection{Accountability}
Traditional AI-based security frameworks can neither entrust individuals with the responsibility of safeguarding and controlling information, nor can hold any individuals accountable to appropriate authorities in the event of loss or misuse of information. It's tough to identify a single individual responsible for a breach. On the other hand, XAI assures accountability by associating each activity with a specific individual rather than a group or an ID. This compels security frameworks to become more resistant to cyber attacks.

\begin{figure*}[h!]
\centering
\includegraphics[width=.90\textwidth]{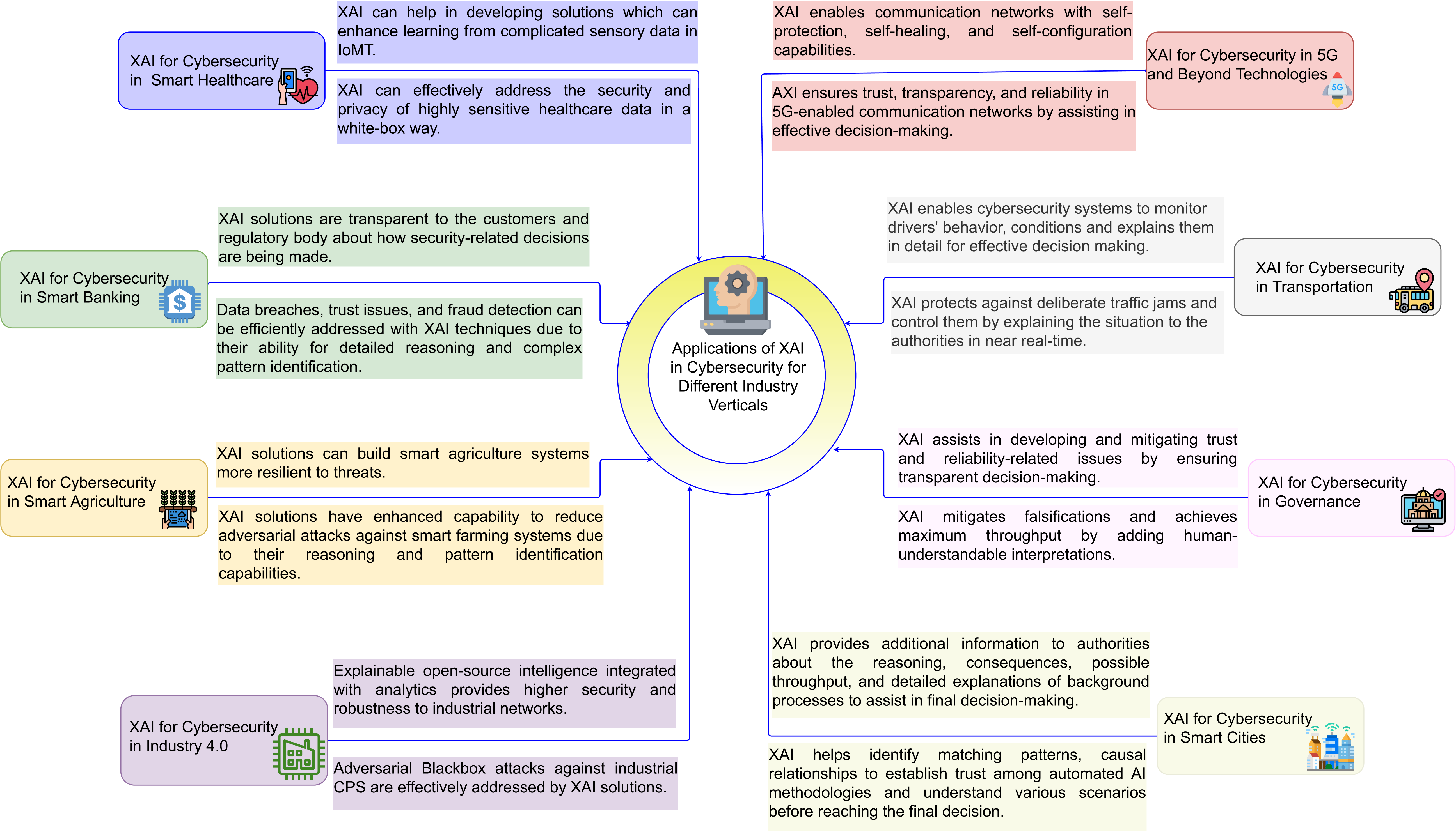}
\caption{Applications of XAI for cybersecurity}
\label{XAIapps}
\end{figure*}

\section{Applications of XAI in Cybersecurity for Different Industry Verticals}

This section provides a comprehensive overview of the potentials of XAI for cybersecurity in distinct industries including smart healthcare, smart banking, smart agriculture, Industry 4.0, smart cities,  smart governance, smart transportation and, 5G and beyond technologies. For each of the applications, it firstly explains the motivation describing the importance of cybersecurity for the use case. Then, it analyzes the requirement of AI-enabled cybersecurity in the domain. Furthermore, it also introduces some important existing works under each of the applications. Lastly, it explains how XAI can enhance the cybersecurity for each application. Fig. ~\ref{XAIapps} depicts various applications of XAI for cybersecurity. 

\subsection{XAI for Cybersecurity in Smart Healthcare}
1) Motivation: Smart healthcare systems have been paving the way into the healthcare industry since last few years. This advanced intelligent system integrates wearable devices, internet-of-medical-things (IoMT), health applications and internet to access sensitive data of different stakeholders and attempts to ensure efficient service with optimized resources. Even though smart healthcare is the future of healthcare industry, the issue of cybersecurity in this industry has been still overlooked \cite{han2021}. One of the motivations for the attackers is that the healthcare data is highly sensitive than any other data. Attackers can obtain health service or prescribed medicines by stealing patients' identities. Filing fraudulent claims and earning huge amount of money is another motivating factor for such malicious people. Attackers can even steal information when some medical institutions keep sufficient information to open fraudulent bank accounts or to secure loans. Apart from stealing information, attackers can deploy ransomware to encrypt sensitive information and, in return, gain huge amount of ransom \cite{kruse2017}\cite{coventry2018}. Moreover, IoMT manufacturers focus more on design features than the all important security aspect. As a result, confidentiality, integrity and availability (CIA) triad related issues are rising prominently \cite{nasiri2019}.

2) Requirements: The healthcare data are not only having diversity but also are in huge volume and great velocity. 
In recent past, AI solutions have been proven more secure, economical and reliable for healthcare industry. These data-driven AI techniques can: (a) process structured, unstructured, semi-structured data; (b) process sequential and multi-dimensional data; (c) provide improved accuracy; (d) provide extremely low latency. As a result, AI-based algorithms are widely used for clinical decision-making, image-based disease diagnosis, classification of clinical documents, online surgery and many other healthcare applications. Due to the aforementioned characteristics of AI-based techniques, they are exploited to be employed for automated detection and protection from different privacy issues and threats, and for safeguarding the sensitive data of the  stakeholders. Furthermore, AI-enabled solutions integrated with big data analytics has been used to address complex cybersecurity threats such as phising attacks, medjacking along with insider threats \cite{wang2021}. \cite{turransky2022}. Some of the recent state-of-the-art on the AI solutions to secure healthcare applications are discussed here. Abie et. al \cite{abie2019} conceptualized an AI-based framework for CPS-IoT to simulate human cognitive behavior for foreseeing and reacting to novel security and privacy threats. A formal modeling-based framework to model and investigate the security of smart healthcare systems was proposed by Haque et. al \cite{haque2021} which uses ML to gain system knowledge in order to create potential attack vectors. Kelli et. al \cite{kelli2021} proposed a blockchain-enabled four layer ML framework, focusing on access control and protection, to enhance security of healthcare applications with anomaly detection methods, reputation techniques along with visual analytics. 

While AI is transforming healthcare with groundbreaking innovations, it should address the following concerns to enhance healthcare security: (a) It is imperative to learn complex patterns from the raw data gathered from wearable devices; (b) It is vital to develop efficient ML-based automated solutions for cloud-based healthcare systems which address privacy and security of highly sensitive medical data;(c) It is critical to produce more XAI security models, while maintaining prediction accuracy in the medical systems.

3) How XAI can help: As AI models fail to explain the decision-making process due to their black-box operations, it is imperative to develop solutions that render AI outcomes interpretable to healthcare workers. XAI integrated with frameworks such as SHAP, LIME, ELI5, Skater, DALEX, and ALE can assist in developing solutions that can enhance learning from complicated sensory data in IoMT. Furthermore, XAI can effectively address the security and privacy of highly sensitive healthcare data in a white-box way. In addition, the involvement of healthcare workers in XAI medical solutions can produce highly accurate predictions.

\subsection{XAI for Cybersecurity in Smart Banking}

1) Motivation: Smart banking systems are gaining their popularity in the domain of finance. It is completely transforming the finance industry with intelligent decisions, automated processes and personalized services and systems. On one side, these smart banking systems provide ease of operations and efficient banking to the stakeholders, on the other side, the risk of cyber attacks against these systems is rising exponentially \cite{maiti2021}. Money has always been the prime motivation of attackers and therefore, banking systems have been the most preferred target for them. In addition, stealing personal data of customers and account holders is a major security threat for the victims. Moreover, with the increasing popularity of internet and mobile banking, attacks through social engineering have also increased in a big way \cite{chaimaa2021}. Thus, it has become increasingly important to minimize the likelihood of frauds and scams.

2) Requirements: AI techniques provide higher efficiency and security to the banking data having high volume and velocity. Due to the inherent capabilities of AI-based algorithms, they are most suitable for customer classification, lending/credit assessments, auditing, regulatory compliance, investment plan suggestions and many other services. Furthermore, AI-based security intelligent models are useful in building novel customized banking products and services. AI solutions have capabilities to preserve privacy of sensitive banking data. Moreover, AI-based security solutions banking applications can detect malicious patterns and money laundering, and thereby, have potential to enhance detection of contemporary frauds. Saker et. al \cite{sarker2021} presented that AI can be effectively used in building a cybersecurity expert system which can be divided into inference engine and the knowledge based security rules. Demirkan et. al \cite{demirkan2020} recommended the integration of AI with blockchain technology in order to build secure and reliable financial products and enhanced quality-of-services (QoS).

Even though we already witnessed the potential of AI in banking sector, it still needs to address the following security issues: (a) AI automated security needs next level of decision-making process for banking sector; (b) In spite of many AI-based solutions, data breach is still a major concern; (c) A cutting-edge solution is mandatory to provide transparency to the stakeholders for security related banking decisions.

3)  How  XAI  can  help:  As AI techniques are not transparent to the end-users, it is vital to make them transparent in order to provide more fruitful decisions and services. Due to the interpretability and explainability characteristics of XAI models, XAI integrated with frameworks such as with frameworks SHAP, LIME, ELI5, Skater, DALEX, and ALE will provide updated and transparent information to the customers and regulatory bodies about how security-related decisions are being made. Additionally, data breaches, trust issues and fraud detection can be efficiently addressed with XAI techniques due to their ability for detailed reasoning and complex pattern identification. 

\subsection{XAI for Cybersecurity in Smart Agriculture}

1) Motivation: Smart agriculture represents the application of advanced technologies such as IoT, robots, drones, sensors and location systems in agriculture and farming. Smart farming is driven by cloud computing, big data, AI along with augmented reality. Even though smart agriculture is gaining popularity, it faces some critical security challenges in data transparency, traceability and anonymity \cite{vangala2021}. The smart agriculture systems are prone to a variety of attacks such as replay attack and MITM attacks. In addition, security attacks against authentication are quite frequent. Current security measures for IoT devices used in smart farming are inefficient and incompatible. These systems require high security with low computation and low communication costs \cite{ferrag2021a}.

2) Requirements: Agriculture related data received from different heterogeneous devices and sensors have diversity and huge volume. AI techniques have proven to be very effective in detecting crop disease, detecting leaf water stress, assessing soil condition, analyzing behavior patterns, recommending optimal gap between plants, suggesting cultivation preparation and more. Due to the technological innovations, it has become possible to minimize human interference. In recent years, AI-based solutions have shown their potential in smart agriculture against distinct security threats such as data attacks, network and equipment attacks, supply chain attacks and cloud computing attacks \cite{gupta2020}. Some of the recent state-of-the-art on the AI solutions to secure smart agriculture applications are discussed here. Ferrag et. al \cite{ferrag2021} proposed a deep learning-based IDS for mitigating distinct types of DDoS attacks in smart agriculture systems. Udendhran et. al \cite{udendhran2021} derived a secure deep learning architecture for smart agricultural systems where a party can share and receive the private data with untrusted third party using private multiparty ML model.

Even though AI is a handy technique in minimizing human interference and optimizing the decisions, it is imperative to address the following security challenges: (a) The supply chain in smart agriculture needs higher trustworthiness and compliance; (b) Privacy of farming data still remains a major concern; (c) Addressing security in the farming applications employed with 5G and beyond technologies would be a key challenge; (d) Addressing security issues concerned with adversarial ML has been a big challenge.

3) How XAI can help: The integration of XAI with smart agriculture can address the limitations posed by AI solutions. XAI integrated solutions with frameworks such as with frameworks SHAP, LIME, ELI5, Skater, DALEX, and ALE can build smart agriculture systems more resilient to various security threats. Furthermore, the transparency and interpretability of XAI make the farming system trustworthy to the internal and external stakeholders. In addition, XAI solutions have enhanced capability to reduce adversarial attacks against the smart farming systems due to their reasoning and pattern identification capabilities. 

\subsection{XAI for Cybersecurity in Industry 4.0}

1) Motivation: In recent years, Industry 4.0 has significantly changed the way in which factories and workplaces are functioning. Industrial IoT and smart manufacturing have brought some fundamental changes in the way industrial world operates with cyber physical systems (CPSs) \cite{jhaveri2021}. The industries adopting smart factory settings have experienced safer, flexible and more efficient manufacturing. However, securing smart industrial network from conventional and unconventional attacks has been a daunting task. The attackers are always looking to find vulnerabilities high-profile manufacturing systems \cite{manogaran2021}. The attackers may inject malicious add-ins in the industrial systems. Moreover, industrial IoT devices are prone to adversarial Trojan. Along with these attacks, data theft, device hijacking, jamming, forging, MITM, DoS/DDoS and stealthy attacks are major threats for industrial control systems, programmable logic controllers, SCADA systems and human machine interfaces \cite{alcaraz2019} \cite{sengupta2020}. In order to avoid serious security breaches and to build trustworthy environment for future small-scale industries, it is vital to keep the security measures intact in industries. 

2) Requirements: AI techniques have potential to solve complex engineering problems and to simulate complicated industrial operations. AI algorithms are extremely useful in decision-making of manufacturing operations and supply chain tasks. In addition, they are very useful in predicting future failures, reducing cycle duration, enhancing process operations, improving product quality, controlling processes, solving real-time issues and optimizing manufacturing operations. Due to the inherent capabilities, they improve overall manufacturing process and supply chain, enhance product quality, reduce production costs and improve customer satisfaction. Rahman et. al \cite{rahman2022} proposed a sustainable and secure AI-based data protection scheme to efficiently detect advanced persistent threats (APT) at the edge and to record the APT detection history in a blockchain ledger. Lv et. al \cite{lv2020} designed an AI-based automated real-time mechanism to improve prediction, accuracy and response time against a variety of information network threats in industrial networks. 

While AI has already shown its promising applications in industries, it still faces several security challenges in the Industry 4.0 landscape: (a) Due to AI's 'Blackbox' characteristic, it is a big security risk to use it as a standalone unit in critical industrial applications; (b) Industrial automation requires tremendous amount of trust and transparency which AI is unable to provide; (c) Security analysts find it extremely difficult to leverage AI-based decisions for industrial control systems.   

3) How XAI can help: XAI helps security personnel by addressing the transparency and explainability issues of AI. Fusion of XAI with frameworks such as SHAP, LIME, ELI5, Skater, DALEX, and ALE, along with conventional threat intelligence techniques can be used to address a variety of cyber threats on industrial CPSs. Explainable open-source intelligence integrated with analytics provides higher security and robustness to industrial networks. Furthermore, adversarial blackbox attacks against industrial CPS can be effectively addressed by XAI solutions.

\subsection{XAI for Cybersecurity in Smart Cities}

 \subsubsection{Motivation} 
 Smart cities applications are transforming the lives of individuals in numerous verticals such as smart healthcare, smart energy, smart transportation, smart retail, smart environment, smart automotive and many more. On one end, the smart cities applications are sophisticating the lives of the citizens of the society, on the other end, these applications are also suffering from privacy and security-related issues. Most of the smart cities applications are integrated with information and communications technologies such as IoT, blockchain and AI, to handle smart cities’ infrastructures, assets and resources effectively. Such technologies also offer potential advantages to smart cities applications such as providing real-time updates related to electricity faults, efficient energy management using concepts such as smart metering, smart traffic management and others. However, it is a challenging task to make cyber-resilient smart cities infrastructure to protect smart city applications from the threat of cybersecurity attacks. Furthermore, it is a cumbersome task to choose an appropriate cyber-resilient algorithm for smart cities applications due to the possibility of numerous cybersecurity attacks\cite{ahmad2022developing}.

\subsubsection{Requirements} 
AI methodologies have been the driving factor for making data-driven decisions for smart cities applications in various verticals such as smart healthcare, smart transportation, smart agriculture, smart governance, smart industries and many more. However, the major concern with the traditional AI methodologies has been the blockbox-like uninterpretable characteristic while making critical data-driven decisions. In recent times, achieving trust and reliability is very essential to protect smart city infrastructure against any kind of cyber security threats or attack. The key factor in ensuring the security of smart cities applications is the control over explainability and interpretability of automated smart cities applications and almost converting the blackbox nature of AI methodologies into whitebox by integrating XAI into cybersecurity-enabled automated smart cities applications\cite{lund2017smart}. \\

Even though the importance of AI in smart cities applications is widely known to researchers, still, due to its black-box like characteristics, AI models have to deal with numerous privacy and security concerns: (a) Security attacks on traffic signals to manipulate traffic directions; (b) GPS spoofing attacks to create a blackout scenario in the city; (c) Privacy and security attacks on CCTV cameras to hide criminal-like individuals\cite{he2018imminent}.
   
\subsubsection{How XAI can help} 
 The opaqueness of AI methodologies is raising security and trust-related limitations in existing smart cities applications. Therefore, the integration of XAI into IoT and AI-enabled smart cities applications can assist in resolving X-ray black-box model issues and add interpretability and explainability aspects for taking effective data-driven decisions for smart cities applications. However, integrating XAI into existing smart cities infrastructures also raises some requirements which need to be fulfilled to achieve better throughput, trust and cyber-resilience in automated smart cities applications. In scenarios such as drifting and poor performance of data-driven automated smart cities applications, it is important to carry out a detailed analysis of historical data and find similar and dissimilar patterns, and causal relationships to assure robustness and reliability of AI models \cite{zolanvari2021trust}. The explainability and interpretability aspects are key for identifying matching patterns, causal relationships to establish trust among automated AI methodologies and understand various scenarios before reaching the final decision. In general, most smart cities applications operate in high-risk environments. Thus, it is essential to form an appropriate XAI procedure to provide additional information to authorities about the reasoning, consequences, possible throughput and detailed explanation of background processes to assist in final decision making. To provide the discussed additional information to decision making, the AI and IoT systems should be integrated with various methods such as SHAP, LIME, ELI5, Skater, DALEX, and ALE can assist security personnel by addressing the transparency and explainability issues of AI. 

\subsection{XAI for Cybersecurity in Smart Governance}
  \subsubsection{Motivation} 
 The concept of pervasive and ubiquitous computing has envisioned 24x7 connectivity and periodic real-time updates. However, due to enormous connectivity starting from a coffee mug, grocery store to gas stations, certain cybersecurity concerns have been raised by attackers, criminals and terrorists. In recent times, cybersecurity has become an unavoidable hurdle for government organizations \cite{timmers2019ethics}. Cybersecurity attacks are deliberate attacks done on various government and private organizations for gaining financial benefits \cite{fardnote}. Another motivation behind attacking government systems using a variety of privacy and security attacks is to access confidential information of the citizens of the country and misuse the sensitive information for misleading purposes such as unauthorized access of an individual's bank account, manipulation of personal details, getting undue advantages over competitors and others \cite{wilson2022sustainable}. Recently, some attackers and terrorists have targeted government agencies to disrupt their routine operations. Such people attempt to target their websites and data centers to create hurdles and slow the implementation pace of crucial projects related to defense, healthcare and safety of the citizens of the country\cite{fardnote}. 
 \subsubsection{Requirements} 
The integration of AI in existing cybersecurity structures can provide several benefits: AI-enabled cybersecurity systems can identify unknown threats such as malware, Trojans and viruses, and can protect the organization against a variety of security attacks. Furthermore, an AI-enabled cybersecurity system can handle large volumes of data, authorization and authentication of users. Such systems can help in reducing response time and early detection of threats \cite{zhang2021artificial}. \\
In recent times, fellow researchers have proposed several AI-enabled cybersecurity approaches. However, the enormous connectivity between various edge devices, data centers, government servers has also exposed the vulnerabilities of AI-enabled cybersecurity systems. The attackers, criminals and terrorists have exploited the weaknesses of existing AI-driven cybersecurity methodologies. Due to the X-ray nature of AI methodologies, it is a cumbersome task to understand the decision-making process carried out by AI-enabled cybersecurity systems. Furthermore, due to the strong heterogeneity of cyberspace and the availability of numerous cyber structures, it is a daunting task for AI methodologies to handle the associated high to medium level risks in the current cybersecurity systems \cite{truong2020artificial}. Another concern for the recent AI-enabled cybersecurity systems is the requirement of a high volume of data to achieve the desired accuracy. Sometimes, due to manufacturing-level design flaws, it becomes a challenging task for AI methodologies to ensure secure communication between heterogeneous devices.\\
    Furthermore, the recently proposed AI-enabled cybersecurity systems are suffering from numerous threats: (a) Absence of transparency in the decision making of AI-enabled cybersecurity systems; (b) The presence of manufacturing or design-level vulnerabilities; (c) The requirement of the high-volume of data to achieve higher throughput; (iv) The cybersecurity structure-based adjustment and modification of AI methodologies. 

\subsubsection{How XAI can help} 
    With the advances in AI-enabled cybersecurity approaches and malicious decision-making of cybersecurity systems, there is an immediate need to introduce explainable or interpretable AI methodologies to assist government and private organizations in taking critical decisions. The explainable ability of AI methodologies can also assist in developing and mitigating trust and reliability-related issues by ensuring transparent and smooth decision-making process \cite{kuppa2021adversarial}.  Recently, fellow researchers have proposed some research works to get more insights about AI models, their workings and debugging. To build accurate AI-enabled cybersecurity systems, it is important to design hybrid approaches by combining frameworks such as SHAP, LIME, ELI5, Skater, DALEX, and ALE with existing AI methodologies and cybersecurity structures \cite{zolanvari2021trust, samtani2020trailblazing}. The integration of XAI approaches with existing AI-enabled cybersecurity systems can mitigate falsifications and achieve maximum throughput by adding human-understandable interpretations.  

\subsection{XAI for Cybersecurity in Smart Transportation}
 \subsubsection{Motivation} 
    With the innovations and advances in transportation systems, travel and trade, the world has become a small global village. The latest transportation technologies have eased the process of transporting people and goods in various geographical regions of the world. The advent of the latest technologies such as Vehicular Ad-hoc Networks(VANETs), IIoT, blockchain and software-defined networks(SDNs) have seamlessly integrated various components of the transportation systems, and have also increased the complexity of operations \cite{kelarestaghi2018intelligent}. Due to this, the current smart transportation systems possess potential threats such as traffic jams or disruption of transportation-related financial services such as ticketing or billing. Furthermore, the motivation behind conducting various transportation attacks is to access confidential information of users and manipulate them for misleading purposes, diverting traffic by creating chaos-like situations to transport explosives or illegal intoxicants, harmful chemicals and radioactive materials \cite{caire2017human}. 
    
    \subsubsection{Requirements} 
     AI-enabled intrusion detection systems (IDSs) and intrusion prevention systems(IPSs) can be employed to provide protection against a variety of security attacks to smart transportation systems \cite{andrade2020comprehensive}. Furthermore, the design and development of a pre-trained AI methodology can be done to get information about potential risks and attacks related to smart transportation systems. \\
  Nevertheless, as every coin has two sides, AI-enabled cybersecurity systems use ML methodologies that may have a vulnerability during the training phase which can be exploited by attackers by creating deliberate disruption in the selection process of training data \cite{chan2019survey}. For instance, if noise is inserted with the training data using attacks such as poisoning and envision, the security of the entire transportation system can be compromised. Furthermore, the AI-enabled IDS and IPS systems are at an initial stage of research, and therefore, it is essential to be aware of the impacts of cybersecurity attacks on smart transportation systems. \\
  Recently, a few security vulnerabilities of the current transportation systems have been exploited by attackers: (a) Blocking of a traffic lane to navigate the traffic in the wrong direction; (b) Disruption of transportation-related financial services using DoS and distributed DoS (DDoS) attacks.
     
\subsubsection{How XAI can help} 
The integration of XAI methodologies such as LIME, SHAP, ELI5, Dalex, and Skater with existing AI methodologies and cybersecurity structures have enabled better explainability and interpretation of the particular scenario to assist in the driver decision-making process \cite{abdel2021federated}. The XAI methodologies enabled existing AI-enabled cybersecurity systems to monitor drivers' behavior, conditions, and detailed explanation for effective decision making.  For instance, the XAI integrated systems can detect and explain accident-like situations in advance and ensure the safety of vehicles. The XAI-enabled cybersecurity systems integrated with frameworks such as SHAP, LIME, ELI5, Skater, DALEX, and ALE can also protect against deliberate traffic jams, and control them by explaining the situation to the authorities in real-time. Furthermore, the XAI-enabled cybersecurity systems can detect the alcoholic conditions of a driver, inform the house members and take appropriate decisions. Recently, fellow researchers have proposed the integration of XAI cybersecurity systems with Human-Computer Interface (HCI) methodologies for a better explanation, interpretation, and decision-making \cite{kashef2021smart}. 
     
\subsection{XAI for Cybersecurity in 5G and Beyond Technologies}
 \subsubsection{Motivation} 
 The advent of 5G and beyond technologies has enabled seamless communication between various edge devices, communication networks and cloud servers. The development of the integration of 5G and beyond technologies with the aforementioned latest technologies has enabled the concept of pervasive and ubiquitous computing by connecting heterogeneous devices distributed in large geographical regions \cite{khan2019survey}. However, in recent times, attackers have targeted the 5G and beyond technologies enabled communication networks and exploited their potential vulnerabilities. The main motivation behind attacking the 5G and 6G-enabled systems is to interfere and disrupt the operational infrastructure by inserting malware. The attackers also conduct such attacks to manipulate the traffic of the network or to disconnect access of resources connected to 5G-enabled systems \cite{ahmad2018overview}. Furthermore, the 5G-enabled networks makes use of the programmable architectures such as Virtual Network Functions(VNFs) and SDNs. Recently, attackers have started attacking such programmable architectures by adding malicious VNFs in the network to stop users from accessing dedicated resources. 
 
 \subsubsection{Requirements} 
 The integration of AI methodologies with existing cybersecurity structures can assist the 5G-enabled IIoT systems in detecting the threats and possibilities of security attacks in advance. The AI-enabled cybersecurity frameworks are also capable of analysing and diagnosing the communication network for the identification of various risks such as malware, viruses or Trojans \cite{kuzlu2021role}. \\ 
 AI-enabled cybersecurity systems integrated with 5G and beyond technologies incur enormous cost of development and computing. Currently, attackers are also using AI-based viruses and systems to counter the protection provided by AI-enabled cybersecurity systems. It is a challenging task for AI-integrated cybersecurity frameworks integrated with 5G and beyond technologies to detect unknown threats. The AI-integrated cybersecurity systems can only detect threats based on the trained datasets. Moreover, in IIoT environments, due to the integration of many heterogeneous hardware, it is almost impossible for an AI-enabled cybersecurity framework to ensure complete security due to compatibility and platform diversity issues \cite{sarker2021ai}.\\
 In recent times, attackers have launched several attacks on 5G and AI-enabled cybersecurity systems : (a) Manipulation of SDNs for disconnecting users from accessing dedicated resources;  (b) Use of malicious VNFs to disrupt the operations of IIoT infrastructures; (c) Due to the unavailability of 5G and beyond technology security standards for IIoT sytems, attackers have exploited the design flaws and configurations of communication networks.   
 
 \subsubsection{How XAI can help} 
 The integration of XAI-enabled cybersecurity systems with 5G and beyond technologies can enable communication networks with self-protection, self-healing, and self-configuration capabilities. The distributed ML methodologies can monitor and analyze the 5G-enabled communication network for security threats, can conduct real-time action, and provide better explainability of the threat scenario to the users \cite{rawal2021recent}. The integration of XAI methodologies with frameworks such as SHAP, LIME, ELI5, Skater, DALEX, ALE, AI-enabled IDSs, and IPSs can deal with a variety of security attacks such as poisoning, MITM, evasion, API-based attacks, DoS, DDoS, and many more. They can also overcome the limitations of existing AI-enabled cybersecurity frameworks by providing better interpretation and explanation of the network threat scenarios \cite{mahbooba2021explainable}. Furthermore, the integration of XAI models with existing AI-enabled cybersecurity structures ensures trust, transparency, and reliability in 5G-enabled communication networks by assisting in effective decision-making.  

\begin{table*}[h!]
\centering
\caption{{Summary of applications}}
\label{tab:Applications}
\resizebox{\textwidth}{!}{%
\begin{tabular}{|c|p{10cm}|
>{\columncolor[HTML]{FFFFC7}}c 
>{\columncolor[HTML]{FFFFC7}}c 
>{\columncolor[HTML]{FFFFC7}}c 
>{\columncolor[HTML]{FFFFC7}}c 
>{\columncolor[HTML]{FFFFC7}}c 
>{\columncolor[HTML]{FFFFC7}}c 
>{\columncolor[HTML]{FFFFC7}}c
>{\columncolor[HTML]{FFFFC7}}c |
>{\columncolor[HTML]{C0F3C0}}c 
>{\columncolor[HTML]{C0F3C0}}c 
>{\columncolor[HTML]{C0F3C0}}c 
>{\columncolor[HTML]{C0F3C0}}c 
>{\columncolor[HTML]{C0F3C0}}c 
>{\columncolor[HTML]{C0F3C0}}c |}
\hline
 &
   &
  \multicolumn{8}{c|}{\cellcolor[HTML]{FFFFC7}Application} &
  \multicolumn{6}{c|}{\cellcolor[HTML]{C0F3C0}Technical Aspect} \\ \cline{3-16} 
{Ref.No.} &
  {Description} &
  \multicolumn{1}{c|}{\cellcolor[HTML]{FFFFC7}{\rotatebox[origin=c]{90}{~Smart Healthcare ~}}} &
  \multicolumn{1}{c|}{\cellcolor[HTML]{FFFFC7}{\rotatebox[origin=c]{90}{~Smart Banking~}}} &
  \multicolumn{1}{c|}{\cellcolor[HTML]{FFFFC7}{\rotatebox[origin=c]{90}{~Smart Agriculture~}}} &
  \multicolumn{1}{c|}{\cellcolor[HTML]{FFFFC7}{\rotatebox[origin=c]{90}{~Industry 4.0~}}} &
  \multicolumn{1}{c|}{\cellcolor[HTML]{FFFFC7}{\rotatebox[origin=c]{90}{~Smart Cities~}}} &
  \multicolumn{1}{c|}{\cellcolor[HTML]{FFFFC7}{\rotatebox[origin=c]{90}{~Smart Governance~}}} &
  \multicolumn{1}{c|}{\cellcolor[HTML]{FFFFC7}{\rotatebox[origin=c]{90}{~Smart Transportation~}}} &
  {\rotatebox[origin=c]{90}{~~}} &
  \multicolumn{1}{c|}{\cellcolor[HTML]{C0F3C0}{\rotatebox[origin=c]{90}{~Confidentiality~}}} &
  \multicolumn{1}{c|}{\cellcolor[HTML]{C0F3C0}{\rotatebox[origin=c]{90}{~Integrity~}}} &
  \multicolumn{1}{c|}{\cellcolor[HTML]{C0F3C0}{\rotatebox[origin=c]{90}{~Availability~}}} &
  \multicolumn{1}{c|}{\cellcolor[HTML]{C0F3C0}{\rotatebox[origin=c]{90}{~Privacy~}}} &
  \multicolumn{1}{c|}{\cellcolor[HTML]{C0F3C0}{\rotatebox[origin=c]{90}{~Accuracy/Reliability~}}} &
  \multicolumn{1}{c|}{\cellcolor[HTML]{C0F3C0}{\rotatebox[origin=c]{90}{~Response Time/QoS~}}} 
   \\ \hline
 \cite{haque2021}&
   Proposed a formal modeling-based framework to model and investigate the security of smart healthcare systems &
  \multicolumn{1}{c|}{\cellcolor[HTML]{FFFFC7}$\checkmark$} &
  \multicolumn{1}{c|}{\cellcolor[HTML]{FFFFC7}} &
  \multicolumn{1}{c|}{\cellcolor[HTML]{FFFFC7}} &
  \multicolumn{1}{c|}{\cellcolor[HTML]{FFFFC7}} &
  \multicolumn{1}{c|}{\cellcolor[HTML]{FFFFC7}} &
  \multicolumn{1}{c|}{\cellcolor[HTML]{FFFFC7}} &
  \multicolumn{1}{c|}{\cellcolor[HTML]{FFFFC7}} &
   &
  \multicolumn{1}{c|}{\cellcolor[HTML]{C0F3C0}} &
  \multicolumn{1}{c|}{\cellcolor[HTML]{C0F3C0}} &
  \multicolumn{1}{c|}{\cellcolor[HTML]{C0F3C0}} &
  \multicolumn{1}{c|}{\cellcolor[HTML]{C0F3C0}} & 
  \multicolumn{1}{c|}{\cellcolor[HTML]{C0F3C0}$\checkmark$} & 
   \\ \hline
   
    \cite{kelli2021}&
   Presented a blockchain-enabled four layer machine learning framework to enhance data protection, anomaly detection and trust &
  \multicolumn{1}{c|}{\cellcolor[HTML]{FFFFC7}$\checkmark$} &
  \multicolumn{1}{c|}{\cellcolor[HTML]{FFFFC7}} &
  \multicolumn{1}{c|}{\cellcolor[HTML]{FFFFC7}} &
  \multicolumn{1}{c|}{\cellcolor[HTML]{FFFFC7}} &
  \multicolumn{1}{c|}{\cellcolor[HTML]{FFFFC7}} & 
  \multicolumn{1}{c|}{\cellcolor[HTML]{FFFFC7}} &
  \multicolumn{1}{c|}{\cellcolor[HTML]{FFFFC7}} &
   &
  
  \multicolumn{1}{c|}{\cellcolor[HTML]{C0F3C0}$\checkmark$} & 
  \multicolumn{1}{c|}{\cellcolor[HTML]{C0F3C0}} &
  \multicolumn{1}{c|}{\cellcolor[HTML]{C0F3C0}} &
  \multicolumn{1}{c|}{\cellcolor[HTML]{C0F3C0}} & 
  \multicolumn{1}{c|}{\cellcolor[HTML]{C0F3C0}$\checkmark$} & 
   \\ \hline
   
    \cite{sarker2021}&
   Presented a cybersecurity expert system containing inference engine and the knowledge based security rule &
  \multicolumn{1}{c|}{\cellcolor[HTML]{FFFFC7}} &
  \multicolumn{1}{c|}{\cellcolor[HTML]{FFFFC7}$\checkmark$} &
  \multicolumn{1}{c|}{\cellcolor[HTML]{FFFFC7}} &
  \multicolumn{1}{c|}{\cellcolor[HTML]{FFFFC7}} &
  \multicolumn{1}{c|}{\cellcolor[HTML]{FFFFC7}} &
  \multicolumn{1}{c|}{\cellcolor[HTML]{FFFFC7}} &
  \multicolumn{1}{c|}{\cellcolor[HTML]{FFFFC7}} &
   &
  \multicolumn{1}{c|}{\cellcolor[HTML]{C0F3C0}} & 
  \multicolumn{1}{c|}{\cellcolor[HTML]{C0F3C0}} &
  \multicolumn{1}{c|}{\cellcolor[HTML]{C0F3C0}} &
  \multicolumn{1}{c|}{\cellcolor[HTML]{C0F3C0}} & 
  \multicolumn{1}{c|}{\cellcolor[HTML]{C0F3C0}$\checkmark$} & 
   \\ \hline
   
 \cite{demirkan2020}&
   Proposed integration of AI with blockchain technology in order to build secure and reliable financial product &
  \multicolumn{1}{c|}{\cellcolor[HTML]{FFFFC7}} &
  \multicolumn{1}{c|}{\cellcolor[HTML]{FFFFC7}$\checkmark$} & 
  \multicolumn{1}{c|}{\cellcolor[HTML]{FFFFC7}} &
  \multicolumn{1}{c|}{\cellcolor[HTML]{FFFFC7}} &
  \multicolumn{1}{c|}{\cellcolor[HTML]{FFFFC7}} &
  \multicolumn{1}{c|}{\cellcolor[HTML]{FFFFC7}} &
  \multicolumn{1}{c|}{\cellcolor[HTML]{FFFFC7}} &
   &
  \multicolumn{1}{c|}{\cellcolor[HTML]{C0F3C0}} & 
  \multicolumn{1}{c|}{\cellcolor[HTML]{C0F3C0}$\checkmark$} &
  \multicolumn{1}{c|}{\cellcolor[HTML]{C0F3C0}} &
  \multicolumn{1}{c|}{\cellcolor[HTML]{C0F3C0}} & 
  \multicolumn{1}{c|}{\cellcolor[HTML]{C0F3C0}$\checkmark$} & $\checkmark$
   \\ \hline
   
    \cite{ferrag2021}&
   Devised a deep learning-based IDS for mitigating distinct types of DDoS attacks  &
  \multicolumn{1}{c|}{\cellcolor[HTML]{FFFFC7}} &
  \multicolumn{1}{c|}{\cellcolor[HTML]{FFFFC7}} &
  \multicolumn{1}{c|}{\cellcolor[HTML]{FFFFC7}$\checkmark$} & 
  \multicolumn{1}{c|}{\cellcolor[HTML]{FFFFC7}} &
  \multicolumn{1}{c|}{\cellcolor[HTML]{FFFFC7}} &
  \multicolumn{1}{c|}{\cellcolor[HTML]{FFFFC7}} &
  \multicolumn{1}{c|}{\cellcolor[HTML]{FFFFC7}} &
   &
  \multicolumn{1}{c|}{\cellcolor[HTML]{C0F3C0}} & 
  \multicolumn{1}{c|}{\cellcolor[HTML]{C0F3C0}} &  
  \multicolumn{1}{c|}{\cellcolor[HTML]{C0F3C0}$\checkmark$} &
  \multicolumn{1}{c|}{\cellcolor[HTML]{C0F3C0}} & 
  \multicolumn{1}{c|}{\cellcolor[HTML]{C0F3C0}} & 
   \\ \hline
   
       \cite{udendhran2021}&
   Devised a secure deep learning architecture for privacy preservation using private multiparty ML model  &
  \multicolumn{1}{c|}{\cellcolor[HTML]{FFFFC7}} &
  \multicolumn{1}{c|}{\cellcolor[HTML]{FFFFC7}} &
  \multicolumn{1}{c|}{\cellcolor[HTML]{FFFFC7}$\checkmark$} & 
  \multicolumn{1}{c|}{\cellcolor[HTML]{FFFFC7}} &
  \multicolumn{1}{c|}{\cellcolor[HTML]{FFFFC7}} &
  \multicolumn{1}{c|}{\cellcolor[HTML]{FFFFC7}} &
  \multicolumn{1}{c|}{\cellcolor[HTML]{FFFFC7}} &
   &
  \multicolumn{1}{c|}{\cellcolor[HTML]{C0F3C0}} & 
  \multicolumn{1}{c|}{\cellcolor[HTML]{C0F3C0}} &  
  \multicolumn{1}{c|}{\cellcolor[HTML]{C0F3C0}} &
  \multicolumn{1}{c|}{\cellcolor[HTML]{C0F3C0}$\checkmark$} & 
  \multicolumn{1}{c|}{\cellcolor[HTML]{C0F3C0}} & 
   \\ \hline
   
    \cite{rahman2022}&
  Presented a secure AI-based data protection scheme to efficiently detect advanced persistent threats &
  \multicolumn{1}{c|}{\cellcolor[HTML]{FFFFC7}} &
  \multicolumn{1}{c|}{\cellcolor[HTML]{FFFFC7}} &
  \multicolumn{1}{c|}{\cellcolor[HTML]{FFFFC7}} & 
  \multicolumn{1}{c|}{\cellcolor[HTML]{FFFFC7}$\checkmark$} &
  \multicolumn{1}{c|}{\cellcolor[HTML]{FFFFC7}} &
  \multicolumn{1}{c|}{\cellcolor[HTML]{FFFFC7}} &
  \multicolumn{1}{c|}{\cellcolor[HTML]{FFFFC7}} &
   &
   
  \multicolumn{1}{c|}{\cellcolor[HTML]{C0F3C0}$\checkmark$} & 
  \multicolumn{1}{c|}{\cellcolor[HTML]{C0F3C0}} &
  \multicolumn{1}{c|}{\cellcolor[HTML]{C0F3C0}} &
  \multicolumn{1}{c|}{\cellcolor[HTML]{C0F3C0}} & 
  \multicolumn{1}{c|}{\cellcolor[HTML]{C0F3C0}$\checkmark$} & 
   \\ \hline
   
   \cite{lv2020}&
   Proposed an AI-based real-time scheme to improve prediction, accuracy and response time against various information network threats in industrial networks &
  \multicolumn{1}{c|}{\cellcolor[HTML]{FFFFC7}} &
  \multicolumn{1}{c|}{\cellcolor[HTML]{FFFFC7}} &
  \multicolumn{1}{c|}{\cellcolor[HTML]{FFFFC7}} & 
  \multicolumn{1}{c|}{\cellcolor[HTML]{FFFFC7}$\checkmark$} &
  \multicolumn{1}{c|}{\cellcolor[HTML]{FFFFC7}} &
  \multicolumn{1}{c|}{\cellcolor[HTML]{FFFFC7}} &
  \multicolumn{1}{c|}{\cellcolor[HTML]{FFFFC7}} &
   &
   
  \multicolumn{1}{c|}{\cellcolor[HTML]{C0F3C0}$\checkmark$} & 
  \multicolumn{1}{c|}{\cellcolor[HTML]{C0F3C0}} &
  \multicolumn{1}{c|}{\cellcolor[HTML]{C0F3C0}$\checkmark$} &
  \multicolumn{1}{c|}{\cellcolor[HTML]{C0F3C0}} & 
  \multicolumn{1}{c|}{\cellcolor[HTML]{C0F3C0}$\checkmark$} & $\checkmark$
   \\ \hline
 \cite{ahmad2022developing}&
 Presented review work on various cyber-resilient methodologies for smart cities applications to provide protection against a variety of security attacks &
  \multicolumn{1}{c|}{\cellcolor[HTML]{FFFFC7}} &
  \multicolumn{1}{c|}{\cellcolor[HTML]{FFFFC7}} &
  \multicolumn{1}{c|}{\cellcolor[HTML]{FFFFC7}} &
  \multicolumn{1}{c|}{\cellcolor[HTML]{FFFFC7}} &
  \multicolumn{1}{c|}{\cellcolor[HTML]{FFFFC7}$\checkmark$} &
  \multicolumn{1}{c|}{\cellcolor[HTML]{FFFFC7}} &
  \multicolumn{1}{c|}{\cellcolor[HTML]{FFFFC7}} &
   &
  \multicolumn{1}{c|}{\cellcolor[HTML]{C0F3C0}$\checkmark$} &
  \multicolumn{1}{c|}{\cellcolor[HTML]{C0F3C0}} &
  \multicolumn{1}{c|}{\cellcolor[HTML]{C0F3C0}} &
  \multicolumn{1}{c|}{\cellcolor[HTML]{C0F3C0}$\checkmark$} &
  \multicolumn{1}{c|}{\cellcolor[HTML]{C0F3C0}} &
   \\ \hline
 
 \cite{zolanvari2021trust}&
 Discussed XAI-enabled cybersecurity framework for securing IIoT systems and environments &
  \multicolumn{1}{c|}{\cellcolor[HTML]{FFFFC7}} &
  \multicolumn{1}{c|}{\cellcolor[HTML]{FFFFC7}} &
  \multicolumn{1}{c|}{\cellcolor[HTML]{FFFFC7}} &
  \multicolumn{1}{c|}{\cellcolor[HTML]{FFFFC7}} &
  \multicolumn{1}{c|}{\cellcolor[HTML]{FFFFC7}$\checkmark$} &
  \multicolumn{1}{c|}{\cellcolor[HTML]{FFFFC7}} &
  \multicolumn{1}{c|}{\cellcolor[HTML]{FFFFC7}} &
   &
  \multicolumn{1}{c|}{\cellcolor[HTML]{C0F3C0}$\checkmark$} &
  \multicolumn{1}{c|}{\cellcolor[HTML]{C0F3C0}} &
  \multicolumn{1}{c|}{\cellcolor[HTML]{C0F3C0}} &
  \multicolumn{1}{c|}{\cellcolor[HTML]{C0F3C0}$\checkmark$} &
  \multicolumn{1}{c|}{\cellcolor[HTML]{C0F3C0}} &
   \\ \hline

 \cite{kuppa2021adversarial}&
 Analyzed the existing AI methodologies, cyber security structures, and XAI-enabled cybersecurity systems &
  \multicolumn{1}{c|}{\cellcolor[HTML]{FFFFC7}} &
  \multicolumn{1}{c|}{\cellcolor[HTML]{FFFFC7}} &
  \multicolumn{1}{c|}{\cellcolor[HTML]{FFFFC7}} &
  \multicolumn{1}{c|}{\cellcolor[HTML]{FFFFC7}} &
  \multicolumn{1}{c|}{\cellcolor[HTML]{FFFFC7}} &
  \multicolumn{1}{c|}{\cellcolor[HTML]{FFFFC7}$\checkmark$} &
  \multicolumn{1}{c|}{\cellcolor[HTML]{FFFFC7}} &
   &
  \multicolumn{1}{c|}{\cellcolor[HTML]{C0F3C0}$\checkmark$} &
  \multicolumn{1}{c|}{\cellcolor[HTML]{C0F3C0}} &
  \multicolumn{1}{c|}{\cellcolor[HTML]{C0F3C0}} &
  \multicolumn{1}{c|}{\cellcolor[HTML]{C0F3C0}} &
  \multicolumn{1}{c|}{\cellcolor[HTML]{C0F3C0}} &
   \\ \hline
   \cite{mahbooba2021explainable}&
 Presented the limitations of 5G-enabled AI methodologies and XAI's potential to overcome and ensure trust and reliability in 5G-enabled communication networks &
  \multicolumn{1}{c|}{\cellcolor[HTML]{FFFFC7}} &
  \multicolumn{1}{c|}{\cellcolor[HTML]{FFFFC7}} &
  \multicolumn{1}{c|}{\cellcolor[HTML]{FFFFC7}} &
  \multicolumn{1}{c|}{\cellcolor[HTML]{FFFFC7}} &
  \multicolumn{1}{c|}{\cellcolor[HTML]{FFFFC7}} &
  \multicolumn{1}{c|}{\cellcolor[HTML]{FFFFC7}} &
  \multicolumn{1}{c|}{\cellcolor[HTML]{FFFFC7}$\checkmark$} &
  &
  \multicolumn{1}{c|}{\cellcolor[HTML]{C0F3C0}$\checkmark$} &
  \multicolumn{1}{c|}{\cellcolor[HTML]{C0F3C0}} &
  \multicolumn{1}{c|}{\cellcolor[HTML]{C0F3C0}$\checkmark$} &
  \multicolumn{1}{c|}{\cellcolor[HTML]{C0F3C0}} &
  \multicolumn{1}{c|}{\cellcolor[HTML]{C0F3C0}$\checkmark$} &
   \\ \hline
\end{tabular}%
}
\end{table*}


%

\section{Research Projects}
In this section, we discuss research projects of XAI across various types of cyber attacks and relevant applications. Table 4 describes research projects on XAI and their relevance in cybersecurity.
\subsection{FeatureCloud}
FeatureCloud is European Union's (EU) funded project. The project was submitted to the H2020-SC1-FA-DTS-2018-2020 call "Trusted digital solutions and Cybersecurity in Health and Care". Finally it was selected for European Union’s Horizon 2020 research and innovation programme under grant agreement number 826078. Implementing XAI in applications such as medical services is highly challenging as it demands both interpretability and security. Feature Cloud is an interdisciplinary project which covers ML, cybersecurity, federated learning, blockchain, XAI and medicine.
The pathology AI is designed and developed based on the feature spaces for medical data that is composed by the FeatureCloud. Also, it applies federated learning approaches in feature spaces which is shared for preserving the privacy of medical data. To facilitate the medical practitioners and other end users, XAI is designed and developed in this project for interpreting the results. The explanation strategy can be chosen based on the user preference. 


\subsection{SPATIAL}
SPATIAL (Security and Privacy Accountable Technology Innovations, Algorithms and machine Learning) is EU funded project. The project was accepted under the Grant agreement number 101021808, with funding of 5 million EUR for EU Horizon 2020. SPATIAL will address the issues faced by blackbox AI and data management in cybersecurity. Key goals of SPATIAL are: (a) Transparency and Explainability: Appropriate methods and techniques to be adapted to ensure AI explainabilty is preserved throughout the security solutions; (b)	Resilience and Privacy: The deliverables are developed in such a way that the system is able to recover from failures in uncontrolled environments as well; (c) Societal Impact for Uptake: To define the organisational polices to ensure uninterrupted trustworthy AI solutions; (d) Education and Skills Building: To build an educational model so as to provide awareness about current and future AI; (e) Transparent AI applications: XAI designed for cyber systems needs to be transparent and understandable between the users and service providers.

\subsection{Blue Hexagon-Hexnet}
Hexnet is the project developed by Blue Hexagon which is the world`s most recognized AI cyber security company. The security operation centre (SOC) analyst is a cybersecurity person responsible for the organisations IT infrastructure and improving the security resilience. The metadata is extracted from the network object and assists the neural net to classify the sample as a malware and non-malware. This simple classification result is not the essential information that the SOC`s might require to understand more about the threat and the reason for it to be treated as malware. The conventional method of assisting SOC is enhanced  by including XAI in the framework. Thereby, SOC could understand and detect the threat and patch the vulnerabilities. Additionally, Hexnet has the potential to map the indicators of compromise (IOC) of that malicious sample to MITRE adversarial Tactics, Techniques and Common Knowledge (ATT\&CK) framework. Hence, Hexnet explains effectively the decisions made by the neural net. Hexnet is resilient to the evasion techniques such as VMWare detection and click detection. Hexnet’s real-time detection and analysis in an organisation with thousands of files and connection will provide reduction of risk and additional cost overhead for large SOC team.


\subsection{LEMNA}
This project was supported in part by NSF grants CNS-1718459, CNS-1750101 and CNS-1717028. Characteristic features of LEMNA are: (a) Augmenting the training data: Since the distribution of contaminated training data might influence the accuracy of the trained model, recent research have proposed various methods to mitigate the misclassifications. The mitigation methods include removing the data points which contribute for misclassifications; (b) Understanding human errors: LEMNA, works by appending the training data to fix under-trained components (instead of removing bad training data). More importantly, LEMNA helps the human analysts to understand these errors before patching them; (c) Complementing robustness: The essential characteristic feature of any model built for security systems needs to resist the adversarial attacks. In recent practice, to build such systems, designers voluntarily introduce adversarial data to the training dataset to make the model more robust. In LEMNA, patching is the method adapted to make the cyber systems more robust. 
(d) Anonymised feature: One remarkable limitation is, LEMNA is not suitable for classifier models for cyber systems built over anonymised feature set. In that case, the model developer needs to map the original feature set with the anonymised feature set to make the LEMNA`s output explainable.

\begin{table*}[h!]
\centering
\caption{Research projects on XAI and their relevance to cybersecurity}
\label{tab:projects}
\begin{tabular}{|
>{\columncolor[HTML]{D4FEFC}}c |
>{\columncolor[HTML]{C8F5C7}}c 
>{\columncolor[HTML]{C8F5C7}}c 
>{\columncolor[HTML]{C8F5C7}}c 
>{\columncolor[HTML]{C8F5C7}}c 
>{\columncolor[HTML]{C8F5C7}}c |
>{\columncolor[HTML]{F7F7E2}}c 
>{\columncolor[HTML]{F7F7E2}}c 
>{\columncolor[HTML]{F7F7E2}}c 
>{\columncolor[HTML]{F7F7E2}}c 
>{\columncolor[HTML]{F7F7E2}}c 
>{\columncolor[HTML]{F7F7E2}}c |}
\hline
\cellcolor[HTML]{ECF4FF} &
  \multicolumn{5}{c|}{\cellcolor[HTML]{ECF4FF}Types of cyber attacks} &
  \multicolumn{6}{c|}{\cellcolor[HTML]{ECF4FF}Relevant applications of XAI in cyber security} \\ \cline{2-12} 
{\cellcolor[HTML]{ECF4FF}Projects (Funding)} &
  \multicolumn{1}{c|}{\cellcolor[HTML]{ECF4FF}{\rotatebox[origin=c]{90}{~Malware~}}} &
  \multicolumn{1}{c|}{\cellcolor[HTML]{ECF4FF}{\rotatebox[origin=c]{90}{~Denial of service~}}} &
  \multicolumn{1}{c|}{\cellcolor[HTML]{ECF4FF}{\rotatebox[origin=c]{90}{~SQL injection~}}} &
  \multicolumn{1}{c|}{\cellcolor[HTML]{ECF4FF}{\rotatebox[origin=c]{90}{~Evasion~}}} &
  \multicolumn{1}{c|}{\cellcolor[HTML]{ECF4FF}{\rotatebox[origin=c]{90}{~Data poisoning~}}} &
  \multicolumn{1}{c|}{\cellcolor[HTML]{ECF4FF}{\rotatebox[origin=c]{90}{~Healthcare~}}} &
  \multicolumn{1}{c|}{\cellcolor[HTML]{ECF4FF}{\rotatebox[origin=c]{90}{~Banking~}}} &
  \multicolumn{1}{c|}{\cellcolor[HTML]{ECF4FF}{\rotatebox[origin=c]{90}{~Industry~}}} &
  \multicolumn{1}{c|}{\cellcolor[HTML]{ECF4FF}{\rotatebox[origin=c]{90}{~Smart City~}}} &
  \multicolumn{1}{c|}{\cellcolor[HTML]{ECF4FF}{\rotatebox[origin=c]{90}{~Governance~}}} &
  \multicolumn{1}{c|}{\cellcolor[HTML]{ECF4FF}{\rotatebox[origin=c]{90}{~Retail~}}} 
   \\ \hline
Feature Cloud (EU) &
  \multicolumn{1}{c|}{\cellcolor[HTML]{C8F5C7}} &
  \multicolumn{1}{c|}{\cellcolor[HTML]{C8F5C7}$\checkmark$} &
  \multicolumn{1}{c|}{\cellcolor[HTML]{C8F5C7}} &
  \multicolumn{1}{c|}{\cellcolor[HTML]{C8F5C7}} &
  $\checkmark$
   &
  \multicolumn{1}{c|}{\cellcolor[HTML]{F7F7E2}$\checkmark$} &
  \multicolumn{1}{c|}{\cellcolor[HTML]{F7F7E2}} &
  \multicolumn{1}{c|}{\cellcolor[HTML]{F7F7E2}} &
  \multicolumn{1}{c|}{\cellcolor[HTML]{F7F7E2}} &
  \multicolumn{1}{c|}{\cellcolor[HTML]{F7F7E2}} &
   \\ \hline
Hexnet (Blue Hexagon) &
  \multicolumn{1}{c|}{\cellcolor[HTML]{C8F5C7}$\checkmark$} &
  \multicolumn{1}{c|}{\cellcolor[HTML]{C8F5C7}} &
  \multicolumn{1}{c|}{\cellcolor[HTML]{C8F5C7}} &
  \multicolumn{1}{c|}{\cellcolor[HTML]{C8F5C7}} &
   &
  \multicolumn{1}{c|}{\cellcolor[HTML]{F7F7E2}$\checkmark$} &
  \multicolumn{1}{c|}{\cellcolor[HTML]{F7F7E2}$\checkmark$} &
  \multicolumn{1}{c|}{\cellcolor[HTML]{F7F7E2}$\checkmark$} &
  \multicolumn{1}{c|}{\cellcolor[HTML]{F7F7E2}} &
  \multicolumn{1}{c|}{\cellcolor[HTML]{F7F7E2}} & $\checkmark$
   \\ \hline
Spatial (EU) &
  \multicolumn{1}{c|}{\cellcolor[HTML]{C8F5C7}} &
  \multicolumn{1}{c|}{\cellcolor[HTML]{C8F5C7}} &
  \multicolumn{1}{c|}{\cellcolor[HTML]{C8F5C7}$\checkmark$} &
  \multicolumn{1}{c|}{\cellcolor[HTML]{C8F5C7}$\checkmark$} &
   $\checkmark$&
  \multicolumn{1}{c|}{\cellcolor[HTML]{F7F7E2}} &
  \multicolumn{1}{c|}{\cellcolor[HTML]{F7F7E2}} &
  \multicolumn{1}{c|}{\cellcolor[HTML]{F7F7E2}} &
  \multicolumn{1}{c|}{\cellcolor[HTML]{F7F7E2}$\checkmark$} &
  \multicolumn{1}{c|}{\cellcolor[HTML]{F7F7E2}$\checkmark$} &
   \\ \hline
Lemna (NSF) &
  \multicolumn{1}{c|}{\cellcolor[HTML]{C8F5C7}$\checkmark$} &
  \multicolumn{1}{c|}{\cellcolor[HTML]{C8F5C7}$\checkmark$} &
  \multicolumn{1}{c|}{\cellcolor[HTML]{C8F5C7}} &
  \multicolumn{1}{c|}{\cellcolor[HTML]{C8F5C7}} &
   $\checkmark$&
  \multicolumn{1}{c|}{\cellcolor[HTML]{F7F7E2}$\checkmark$} &
  \multicolumn{1}{c|}{\cellcolor[HTML]{F7F7E2}$\checkmark$} &
  \multicolumn{1}{c|}{\cellcolor[HTML]{F7F7E2}$\checkmark$} &
  \multicolumn{1}{c|}{\cellcolor[HTML]{F7F7E2}$\checkmark$} &
  \multicolumn{1}{c|}{\cellcolor[HTML]{F7F7E2}$\checkmark$} & $\checkmark$
   \\ \hline
\end{tabular}%
\end{table*}

\section{Lessons Learnt, Challenges, Open Issues and Future Directions}

This section discusses the leanings and lessons learned from surveying the state-of-the-art XAI methodologies related to smart cities applications. In this section, based on the rigorous and detailed review, we have highlighted the research challenges and discussed possible future directions.

\subsection{Security attacks on XAI integrated cybersecurity frameworks}
XAI methodologies protect existing AI-enabled cybersecurity systems from a variety of security attacks by detecting security threats in advance. However, in recent times, attackers have exploited vulnerabilities of XAI integrated cybersecurity frameworks using a variety of attacks \cite{kuppa2021adversarial}: (a) poisoning attacks were used to manipulate or damage the training data; (b) adversarial attacks were applied to evade authentication mechanisms such as XAI-enabled face authentication system; (iii) Counterfactual gaming attacks were used to manipulate defense/offense mechanisms and exploit vulnerabilities of XAI-enabled cybersecurity frameworks.
\\ 
\textbf{Possible Alternatives} A detailed experimental study of the impacts of various attacks on XAI methodologies can be done on a variety of cybersecurity structures and, structured and unstructured data types \cite{kuppa2021adversarial}.

\subsection{Managing balance between security and usability of XAI integrated cybersecurity systems}
Cybersecurity experts need to maintain the trade-off between security and usability aspects of the newly introduced XAI-enabled cybersecurity systems. It is important to consider the fusion of security and usability while conceptualization and design stage of the novel anti-virus systems for ease of use.\\ 

\textbf{Possible Alternatives} To maintain the balance between security and usability aspects, there is an immediate need to introduce tools to analyze and debug the model and its data internals.  \cite{kuppa2021adversarial}.

\subsection{Legal Compliance}
Most of the XAI-enabled cybersecurity frameworks and anti-virus systems are at their initial stage of research. It is possible to misuse interpretation and explainability aspects of XAI models for misleading motives such as breaching XAI-enabled face authentication mechanisms.   \\ 

\textbf{Possible Alternatives} To ensure the safety and security of newly introduced cybersecurity frameworks against attackers, terrorists and wrongdoers, it is essential to formulate appropriate rules and regulations to ensure the constructive use of XAI-enabled cybersecurity systems for the safety of the citizens \cite{ahmed2022artificial}.

\subsection{Performance of XAI Algorithms}
 Even though XAI models look promising, they face several challenges in balancing interpretability and performance aspects. Moreover, security challenges increases when XAI algorithms are employed for decision-making in light-weight nodes such as in IoT systems. Additionally, complex and massive data with high traffic rate is one of the biggest challenges for such systems. Therefore, it is important to address the issues related to optimization of XAI algorithms. Furthermore, in order to employ XAI algorithms in practical security applications, time complexity of the algorithms ought to be addressed. 
 
 \textbf{Possible Alternatives} XAI algorithms can be optimized by considering multiple constraints such as computational resources, model accuracy and computational power. In addition, efficiency of XAI algorithms against big and complex data should be given high consideration. Moreover, accurate decision-making can be targeted by considering data diversity, data rate and data quality \cite{arrieta2020}.

\subsection{Privacy and Data Protection}
XAI models need to consider privacy issues explicitly during system life cycle. It is imperative to respect the right to privacy of each and every person. Along with consideration of data integrity, XAI solutions should ensure the data processing in a way which assures privacy. Data privacy should be protocol driven and indicative of the users and circumstances under which the data can be accessed. Meanwhile, the fusion of the outputs of XAI models can make the overall model non-transparent and non-explainable. In addition, if the model explainability is reversible with knowledge fusion, the differential privacy of the model gets compromised.

\textbf{Possible Alternatives} XAI solutions can explain the loss of privacy with the use of privacy loss and intent loss scores. Moreover, model fusion can become a powerful technique to address the limitations of XAI with post-hoc XAI techniques. Meanwhile, while designing XAI models, it is to be taken care that, the model is not reversible \cite{arrieta2020}. 


\section{Conclusion}
XAI is a framework which helps to enable understanding and interpretations of predictions with the help of ML models. It provides understandability, transparency and justifiability to the results generated by the traditional AI systems. Cybersecurity is one such area where the application of AI helps to analyse dataset and track wide spectrum of security threats and malicious activities. The presented work provides a state-of-the art review of the application of XAI in cybersecurity. At the outset, a brief introduction to cybersecurity is presented highlighting the different types of attacks and its implications. The traditional risk management systems and relevant standards are discussed in details. Although these preventive measures have given promising results but they have associated challenges in terms of their lags in proactive risk management. This set the stage for the implementation of ML algorithms which help to detect attacks and anomalies with enhanced accuracy resembling human behaviour. The ability of AI-based applications in effectively managing the vulnerability of the network by identification of various threats, is  established. However, there still exist associated challenges wherein attacker strategically foil the AI algorithms, making it immune, acting as a vector for various forms of attacks. Additionally, these AI-based systems require sophisticated expertise to develop and maintain, contrarily consuming computational power, data and resources. The results generated are black-boxed wherein understanding and visualization is constrained to input and output having inability to provide traceability between both the parameters. This drawback lays the foundation of XAI which provides opaqueness and explainability to the results leading to accurate and justifiable decision making. 

The article provides an exhaustive description of the implementation of XAI in various verticals namely in smart healthcare, smart banking, smart agriculture, smart cities, smart governance,  smart transportation, Industry 4.0, 5G and beyond technologies explicitly. The contribution of XAI in all of these applications are explored from the perspective of cybersecurity. In continuation to this, the various research projects and potential real-world applications are discussed. As any coin has both sides, XAI implementations also have potential challenges relevant to the applications and projects discussed. These identified challenges are scrutinized and potential alternate solutions are proposed which provide guidance for future direction of research. The paper, thus, presents an exhaustive review of the potential application of XAI and highlights its ability in effective decision making. To conclude, this extensive review directs the need for fusion models in XAI using diversified quality data to visualize further optimized benefits of XAI implementations in real-world.

\bibliographystyle{IEEEtran}
\bibliography{ref}

\end{document}